\documentclass[manuscript, nonacm=true, authorversion=true]{acmart}
\AtBeginDocument{%
  }

\setcopyright{acmlicensed}
\copyrightyear{2026}
\acmYear{2026}
\acmDOI{XXXXXXX.XXXXXXX}
\usepackage{todonotes}
\usepackage{graphicx}
\usepackage{subcaption}
\usepackage{xcolor}
\usepackage[utf8]{inputenc}
\usepackage{newunicodechar}
\usepackage{amsmath}
\newunicodechar{₂}{$_2$}
\usepackage{float}      % in preamble
\usepackage{placeins}
\usepackage{multirow}
\usepackage{booktabs}
\usepackage{multirow}
\usepackage{tabularx}
\usepackage{pifont}     % for check / cross marks
\usepackage{array}      % for column formatting
\renewcommand\footnotetextcopyrightpermission[1]{} 

\begin{document}

\title{Towards a Frugal Photosynthesis Sensing Toolkit for Data-Driven Plant Science Education and Exploration}

 %classroom, Science lab, hobbyiest 
 %democratizing 
% 
\author{Qitong Li}
\email{qitongli2025@u.northwestern.edu}
% \affiliation{%
%   \institution{Northwestern University}
%   \city{Evanston}
%   \state{Illinois}
%   \country{United States}
% }

\author{Raj Nileshbhai Dave}
% \email{rajdave2027@u.northwestern.edu}
% \affiliation{%
%   \institution{Northwestern University}
%   \city{Evanston}
%   \state{Illinois}
%   \country{United States}
% }

\author{Rhema Amanda Phiri}
% \affiliation{%
%   \institution{Northwestern University}
%   \city{Evanston}
%   \state{Illinois}
%   \country{United States}
% }

\author{Leo Zhang}
% \affiliation{%
%   \institution{Northwestern University}
%   \city{Evanston}
%   \state{Illinois}
%   \country{United States}
% }

\author{Xiaoyu Zheng}
% \email{xiaoyuzheng2027@u.northwestern.edu}
% \affiliation{%
%   \institution{Northwestern University}
%   \city{Evanston}
%   \state{Illinois}
%   \country{United States}
% }

\author{Ariana Blake}
% \email{arianablake2028@u.northwestern.edu}
% \affiliation{%
%   \institution{Northwestern University}
%   \city{Evanston}
%   \state{Illinois}
%   \country{United States}
% }

\author{Livia Ford}
% \email{liviaford2028@u.northwestern.edu}
\affiliation{%
  \institution{Northwestern University}
  \city{Evanston}
  \state{Illinois}
  \country{United States}
}

\author{Sarah Jones}
\affiliation{%
  \institution{Chicago Botanic Garden}
  \state{Illinois}
  \country{United States}
}

\author{Susan R. Strickler}
\affiliation{%
  \institution{Chicago Botanic Garden}
  \city{Glencoe}
  \state{Illinois}
  \country{United States}
}
\affiliation{%
  \institution{Northwestern University}
  \city{Evanston}
  \state{Illinois}
  \country{United States}
}

\author{Nivedita Arora}
\affiliation{%
  \institution{Embodied System Lab, Northwestern University}
  \city{Evanston}
  \state{Illinois}
  \country{United States}
}

\renewcommand{\shortauthors}{Li et al.}

\begin{abstract}
Rapid environmental change and advances in data-driven analysis highlight the need not only to use computational tools, but also to foster understanding of the natural world and inspire creativity. Photosynthesis, the process that fuels nearly all life on Earth, provides a compelling context for such learning, particularly in understanding how plants alter their photosynthetic strategies in response to environmental changes. However, existing tools for studying photosynthesis are often inaccessible or limited to demonstrating its presence, rather than capturing its temporal dynamics. We present PhytoBits, a frugal in situ gas-exchange sensing toolkit for distinguishing and teaching photosynthetic strategies. PhytoBits combines leaf enclosure with accessible materials, an off-the-shelf CO\textsubscript{2} sensor, and a low-cost microcontroller, to support multi-day monitoring of plant gas-exchange in educational and research contexts. We validated PhytoBits against research-grade gas-exchange systems, confirming that it identifies C\textsubscript{3} and CAM (Crassulacean Acid Metabolism) photosynthetic pathways. In addition to obligate CAM, PhytoBits also resolves facultative CAM and developmental CAM dynamics in plants. This work presents an early-stage hardware validation; user deployment studies, open-source code dissemination, and automated pathway classification are planned as future work.
\end{abstract}




\maketitle

\begin{figure}[htbp]
  \centering
  \includegraphics[width=0.8\linewidth]{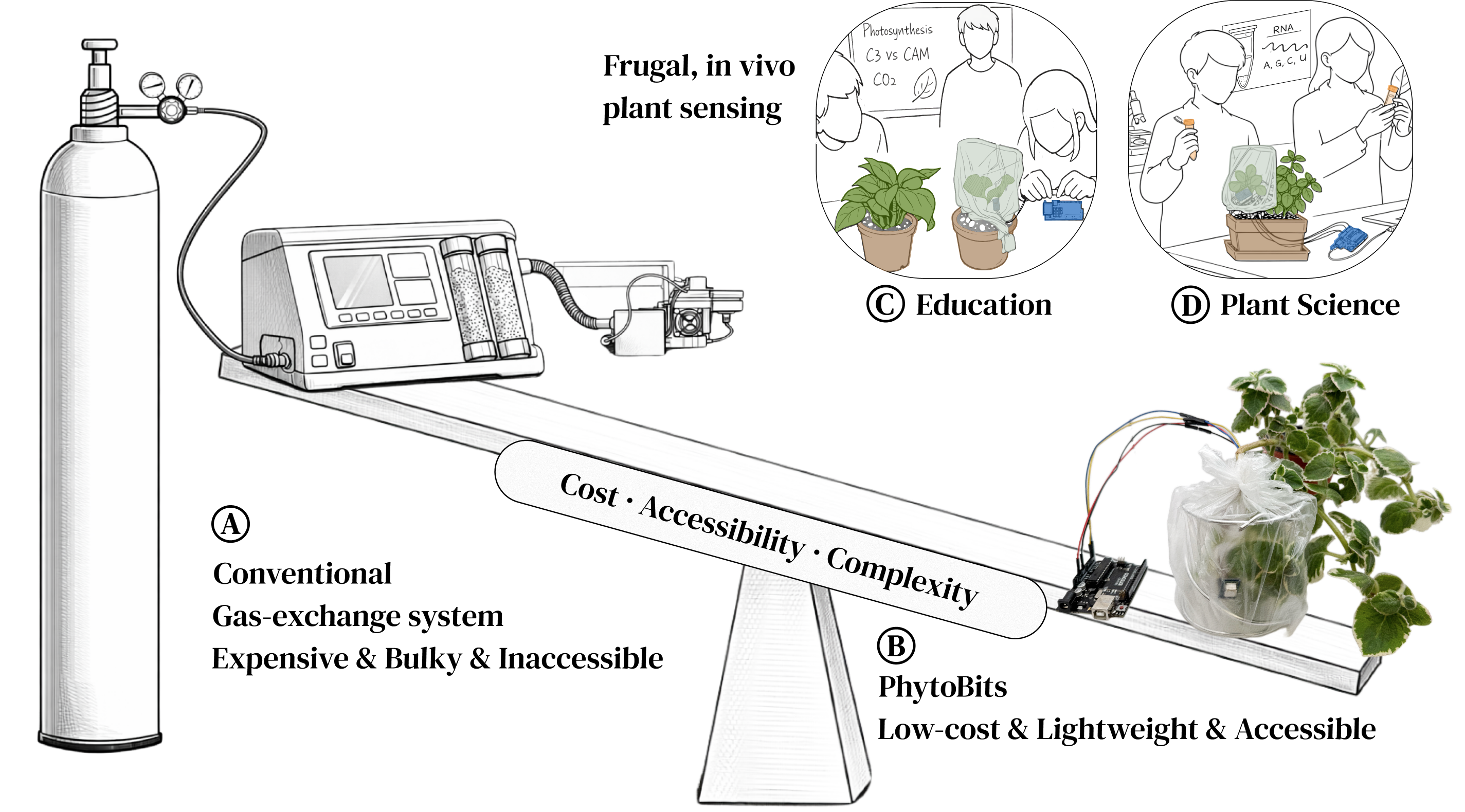}
  \caption{PhytoBits as a frugal plant sensing toolkit, making gas-exchanging plant sensing accessible. (A) A conventional gas-exchange system with CO\textsubscript{2} source and bulky instrumentation, limiting accessibility, (B) PhytoBits toolkit setup for measuring gas-exchange from attached leaves, supporting (C) hands-on photosynthesis education for students and enabling (D) accessible plant science research experiments.
  }
    
  \Description{Comparison between traditional photosynthesis measurement systems and PhytoBits toolkit. Left side shows a LI-COR gas exchange system as a large, expensive laboratory instrument with limited accessibility. Right side shows the PhytoBits toolkit as a compact, low-cost alternative consisting of a microcontroller, CO2 sensor, and lightweight leaf enclosure. The figure illustrates applications in both educational settings (students conducting experiments) and research contexts, demonstrating the toolkit's accessibility and versatility for photosynthesis gas exchange sensing.}
  \label{fig:first_figure}
\end{figure}

\section{Introduction}\label{introduction}
% Motivation
Instruments and tools are extensions of scientific thinking, the capacity to build, and the ability to solve problems \cite{kay1984computer, rheingold1991tools}. For example, the advent of the microscope opened entirely new avenues to study life sciences. Similarly, in the physical computing community, Arduino and Micro:bit have lowered barriers for designers or novice learners with little to no background in computing \cite{badamasi2014working, sentance2017creating}. When designed thoughtfully, tools can lower barriers to entry to new domains, making knowledge accessible and empowering creative pursuits \cite{byagathvalli2021frugal}. 

With this motivation, the PhytoBits toolkit seeks to make the knowledge and metabolic pathways of one of life’s most fundamental processes, photosynthesis, as accessible to students as Arduino made electronics for physical computing. Photosynthesis drives global carbon and water cycles and is essential for understanding stress responses in C\textsubscript{3} crops and succulent species \cite{mirzabaev2023severe}, climate stability, and environmental resilience \cite{Johnson2016Photosynthesis}. In an age where climate change is a critical problem that remains largely ignored in the context of photosynthesis education \cite{photosynthesis_education}, helping students understand photosynthesis is as important as teaching them to code. Studying and understanding this process is necessary for students to meaningfully impact the world. 

There is extensive work in the IoT and ubiquitous computing communities on measuring environmental factors relevant to plants, such as light, soil moisture, and temperature, as indicators of plant health \cite{chandra2022democratizing, nguyen2025portable,coppedeVivoBiosensingBiomimetic2017}. These environmental variables are too indirect and lack a direct connection to \textbf{what is happening inside the plant at a biochemical level during photosynthesis}. For example, soil moisture alone cannot reveal which photosynthetic pathway a plant is using. Photosynthesis is an integrative physiological signal that cannot be inferred from environmental conditions alone. Plants express distinct photosynthetic pathways, C\textsubscript{3}, C\textsubscript{4}, and CAM (Crassulacean Acid Metabolism) \cite{osmond1982functional}, based on their evolutionary adaptations to water and CO\textsubscript{2} availability \cite{keeley2003evolution,winter2019ecophysiology}. Environmental sensing often overlooks the biological richness encoded in photosynthetic metabolism, a dimension that should be made accessible to help students explore, think creatively, and connect science to real-world contexts. 

To capture metabolic biochemical transitions, particularly shifts between photosynthetic modes in response to drought, heat, or growth stage \cite{buschGuidePhotosyntheticGas2024}, plant scientists rely on CO\textsubscript{2} gas-exchange as the gold standard. This approach, however, requires large, bulky, and expensive scientific instrumentation with controlled air flow and CO\textsubscript{2} sources, typically costing on the order of \$10,000 \cite{LICORBiosciences2024,QubitSystems2020,CIDBioScience2024,PPSystems2024}. As a result, such scientific tools remain largely inaccessible to students, educators, and researchers in resource-constrained settings. 
This instrumentation gap is reflected in pedagogy that is mostly limited to explaining photosynthesis with static diagrams or animations, completely limiting the possibility for students to link theory to practice. While some recent commercial educational tools offer portable CO\textsubscript{2} gas sensors paired with wireless data loggers, they fall short of meaningfully capturing or relating measurements to photosynthetic dynamics in practice. They use excised leaves rather than living leaves attached to the plant, and the experiments are over short minutes to hours observation windows, preventing longitudinal in situ study of photosynthetic dynamics \cite{PASCOScientific2026, VernierSoftware2026}. 
Some plant wearables \cite{freyECOvetteUnveilingPlant2025, nassar2018compliant, wu2025wearable} and IoT monitoring systems \cite{dudarov2025influence,nguyen2025portable,saccardi2025low} now measure microclimate, but they essentially focus on environmental sensing and ignore the potential of gas signal dynamics. As a result, they cannot measure stomatal behavior, distinguish C$_3$ from CAM photosynthesis, or reveal metabolic transitions manifesting as changes in gas-exchange timing.
This represents a critical gap: no existing system simultaneously offers affordability, non-invasiveness, multi-day deployment, and the ability to distinguish C\textsubscript{3} and CAM photosynthetic pathways through continuous leaf-level gas-exchange measurements.

Recent advances in frugal science offer a proven alternative model. 
The Foldscope paper microscope, costing less than \$2 to manufacture, has distributed over 2 million units globally, transforming microscopy education in resource-limited regions \cite{cybulski2014foldscope}. This example establishes a compelling precedent: democratization of scientific tools does not require sacrificing scientific utility. Rather, it demands strategic focus on questions where ``good-enough'' quantification suffices to yield actionable biological insight, particularly in educational contexts where the primary goal is conceptual understanding of physiological processes and their environmental drivers \cite{byagathvalli2021frugal, collins2024frugal, chan2021low}.

PhytoBits toolkit is intended to be frugal and pedagogically accessible while retaining adequate reliability for quantitative leaf gas-exchange and environment measurements. 

We contribute:

\begin{itemize}

    \item \textbf{A frugal, flow-free gas-exchange sensing system and reproducible construction process:}
    We introduce a low-cost, no-flow design that replaces the controlled air flow and CO\textsubscript{2} source required by traditional gas-exchange systems. We provide detailed configurations for constructing the toolkit, enabling use in classrooms and research labs.The current design produces semi-quantitative temporal CO\textsubscript{2} and humidity patterns; conversion to absolute leaf flux rates requires per-chamber leak calibration, which is identified as future work.

    \item \textbf{Experimental setups enabling long-term, in situ study of diverse photosynthetic strategies:}
    We demonstrate how the system can capture multi-day gas-exchange signals across C\textsubscript{3} plants and multiple CAM types, including constitutive, facultative, and developmental CAM. The toolkit supports observation of living plants over longer timescales and to study metabolic transitions that are typically inaccessible without professional instruments.
    
    \item \textbf{A toolkit designed to broaden access across classrooms, research labs, and community science:} PhytoBits is designed to serve diverse users across educational, research, and community contexts. In classrooms, students could assemble the toolkit, collect multi-day datasets, and observe how photosynthetic rhythms emerge from authentic biological signals. In research labs, it offers a low-cost alternative for plant screening, metabolic pathway monitoring (e.g., C\textsubscript{3} versus CAM), and generating parallel datasets unattainable with traditional gas-exchange systems. Beyond academia, plant enthusiasts, growers, and citizen scientists can track physiological responses, stress dynamics, and care optimization through quantitative gas-exchange measurements.
    
\end{itemize}

\section{Related Work} \label{section:related_work}

In instructional settings, C\textsubscript{3}, C\textsubscript{4}, and CAM pathways are introduced through static biochemical diagrams, while experimental approaches for observing their physiological signatures remain limited. Distinguishing these pathways in practice depends on resolving diurnal gas-exchange dynamics, yet most classroom and short-term laboratory methods rely on instantaneous measurements that cannot capture day-night rhythms. The difficulty of this observation is evidenced by student attempts to measure differences in short-term laboratory settings, where a lack of 24-hour monitoring and leaf enclosures, prevents the detection of distinct photosynthetic signatures \cite{harris2017spinach}. 
A review of the existing approaches contextualizes the importance of PhytoBits.

\subsection{Photosynthetic modes Measurement Instrumentation and Methods}
Research-grade gas-exchange systems \cite{LICORBiosciences2011, LICORBiosciences2024, CIDBioScience2024, PPSystems2024, QubitSystems2020} offer high precision (down to 0.1-1 parts per million (ppm)) and fast response (minutes-scale)  for instantaneous photosynthesis measurements \cite{huntMeasurementsPhotosynthesisRespiration2003, buschGuidePhotosyntheticGas2024}, but remain inaccessible to most educators and students as equipment costs range from \$10,000 to \$80,000.
Distinguishing photosynthetic pathways from temporal signatures requires multi-day leaf clamping with continuous CO\textsubscript{2} supply, daily chemical changes, and dedicated lab personnel, making such studies far beyond what typical classrooms or field stations can support.

\textbf{Biochemical assays}. Organic acid titration of dawn and dusk leaf sample extractions can detect CAM characteristic malic acid accumulation, but the method is destructive, labor-intensive, and incompatible with tracking the physiological trajectory of a single plant over time \cite{winter2020constitutive, winter2022cam, yang2015comparing}. Stable isotope analysis (\(\delta^{13}\)C via IRMS (Isotope Ratio Mass Spectrometry): \$100,000--150,000 for equipment; \$10--35 per sample) provides a definitive pathway classification but integrates isotopic ratios over weeks to months of growth, missing the dynamic CAM transitions that emerge within days \cite{hubick1989carbon, nogues2008assessing}. Transcriptomics \cite{wai2017temporal, qiu2023mechanisms} and metabolomics \cite{tay2021metabolic} further illuminate mechanistic detail but are prohibitively expensive (\$100--400 per sample), destructive, and disconnected from real-time physiological observation.

\subsection{Emerging Low-Cost Systems for Plant Monitoring}

Recent work has explored low-cost systems for plant and soil monitoring, including frugal photosynthesis rates with air flow control measurements inspired by research-grade gas-exchange systems ~\cite{takaragawa2025establishment, ucschawaii2024par}, automatic gas sampling ~\cite{li2020developing}, and multi-day CO\textsubscript{2} tracking for root or soil respiration ~\cite{dudarov2025influence,nguyen2025portable,saccardi2025low}. These efforts demonstrate how standard CO\textsubscript{2} sensors and open-source hardware can substantially reduce the cost of gas-exchange measurements. Complementary directions in plant sensing have introduced wearable capacitance sensors for water status ~\cite{thalheimerLeafmountedCapacitanceSensor2022}, lightweight cuvettes for volatile emissions ~\cite{freyECOvetteUnveilingPlant2025}, and soft bioelectronic interfaces for in vivo sap and ion monitoring ~\cite{diacciDiurnalVivoXylem2021,gentileBiomimeticBiocompatibleOECT2022}. In parallel, educational platforms focused on soil and environmental monitoring ~\cite{microsoft2025farmbeats,chandra2022democratizing}, CO\textsubscript{2} sensing ~\cite{ribbit2025}, and at-home visualization of plant respiration ~\cite{phytopulse2025} have leveraged inexpensive sensors to broaden access to environmental and plant-science inquiry. Despite this progress, most existing low-cost gas-exchange systems are for short deployments, depend on active airflow and complex enclosures, or emphasize environmental conditions rather than plant-level physiology. As a result, few systems are designed to capture the continuous diurnal gas-exchange patterns of C\textsubscript{3} and CAM photosynthesis in intact, attached leaves.

\subsection{Educational Sensors and Classroom Methods}
Commercial educational platforms (PASCO Wireless CO\textsubscript{2} Sensor, Vernier Go Direct CO\textsubscript{2}) \cite{PASCOScientific2026, VernierSoftware2026} lower financial and operational barriers by bundling relatively affordable hardware (approximately \$200-300) with intuitive data-visualization software \cite{becker2024ecology}. However, these systems are optimized for short-term demonstrations rather than continuous plant-physiology measurements: their metabolism chambers are designed for excised leaves, rendering them unsuitable for observing multi-day photosynthetic dynamics. 

In most classrooms, assays such as floating leaf-disk oxygen evolution \cite{tang2002photosynthetic}, stomatal imprinting \cite{hilu1984convenient}, and chromatography \cite{benson1950path}, engage students in hands-on observation of specific photosynthetic processes or pigment composition within a single laboratory session. Despite their pedagogical value, these activities are destructive, provide short-duration measurements, and lack the continuity needed to reveal pathway-specific rhythms. A single floating leaf-disk measurement reveals instantaneous photosynthetic but cannot show how that rate oscillates across a 24-hour cycle, the temporal signature distinguishing CAM from C\textsubscript{3}, or how plants respond to environmental change.

\subsection{Frugal toolkit design}
Frugal science represents a design paradigm that preserves scientific principles while reducing cost, complexity, and resources, providing accessible alternatives to conventional laboratory equipment and scientific practice \cite{collins2024frugal}. Examples span from paper-based origami microscopes costing less than \$2 \cite{cybulski2014foldscope}, hand-powered centrifuges \cite{bhamla2017hand}, to low-cost CRISPR gene-editing and synthetic biology kits designed for classroom use \cite{collins2024frugal, luo2022demonstrating}. These successes illustrate that, for discovery and education, the precision and expense of research-grade instruments often exceed what is necessary.  

For health applications, smartphone cameras paired with paper calibration cards enable jaundice screening \cite{de2014bilicam}, and the speaker and microphone on phones have been adapted for fetal heart-rate monitoring \cite{garg2025dopfone} and spirometry measurements \cite{Goel:2016:SML:2858036.2858401}. These examples illustrate a core principle of frugal design: by targeting essential signals and adequate measurement resolution, ubiquitous tools can be transformed into previously inaccessible scientific instruments.

PhytoBits extends this philosophy to plant gas-exchange physiology. By prioritizing accessibility and continuous monitoring of intact leaves, PhytoBits adapts frugal science principles to support inquiry-based learning in plant physiology and to broaden participation in hands-on photosynthesis research.

\subsection{Comprehensive Comparison: Methods for Observing C\textsubscript{3} vs. CAM Photosynthesis}

% The following table compares instruments and methods spanning the full spectrum from research-grade (highest precision, highest cost) to educational (accessible, moderate precision) to emerging low-cost systems (accessible, modest precision).

% \newpage
\begin{table}[H]
  \centering
  \scriptsize
  \setlength{\tabcolsep}{3.5pt} % tighten columns to reduce overfull
  \label{tab:comprehensive-comparison}

  \begin{tabularx}{\textwidth}{
    >{\raggedright\arraybackslash}p{0.19\textwidth}
    >{\centering\arraybackslash}p{0.14\textwidth}
    >{\centering\arraybackslash}p{0.29\textwidth}
    >{\centering\arraybackslash}p{0.12\textwidth}
    >{\centering\arraybackslash}p{0.07\textwidth}
    >{\centering\arraybackslash}p{0.11\textwidth}
  }
    \toprule
    \textbf{Instrument} & \textbf{Cost} &
    \textbf{Measurement Duration} & \textbf{Leaf Status} &
    \textbf{Distinguish C\textsubscript{3}/CAM?}$^{\dagger}$ & \textbf{Accessibility} \\
    \midrule

    \multicolumn{6}{c}{\textbf{RESEARCH-GRADE INSTRUMENTS}} \\
    \midrule

    \textbf{LI-COR LI-6400XT/LI-6800} \cite{LICORBiosciences2011, LICORBiosciences2024} & \$40k--80k & continuous monitoring with daily labor & Clamped & \checkmark & Very low \\
    \textbf{CID CI-340} \cite{CIDBioScience2024} & \$19.5k & continuous monitoring with active supervision & Clamped & $\sim$ & Low \\
    \textbf{PP-Systems TARGAS-1} \cite{PPSystems2024} & \$20k--30k & continuous monitoring with daily labor & Clamped & $\sim$ & Low \\
    \textbf{Qubit Q-Box CO650} ~\cite{QubitSystems2020} & \$12.5k & continuous monitoring with active supervision & Clamped & $\sim$ & Low \\
    \midrule

    \multicolumn{6}{c}{\textbf{BIOCHEMICAL ASSAYS}} \\
    \midrule

    \textbf{Organic Acid Titration} \cite{winter2020constitutive, winter2022cam, yang2015comparing} & \$50--200 & 2 discrete time points/day over weeks & Excised (destructive) & \checkmark & Moderate \\
    \textbf{Stable Isotope $\delta$\textsuperscript{13}C (IRMS)}  \cite{hubick1989carbon, nogues2008assessing} & \$10--35/sample + (\$100k+) & Integrated over months (single value per sample) & Excised (destructive) & \checkmark & Very low \\
    \textbf{Transcriptomics / Metabolomics} \cite{wai2017temporal, qiu2023mechanisms,tay2021metabolic} & \$100--400/sample & Discrete snapshots at chosen time points & Excised (destructive) & \checkmark & Very low \\
    \midrule

    \multicolumn{6}{c}{\textbf{EDUCATIONAL TOOLS (COMMERCIAL)}} \\
    \midrule

    \textbf{PASCO Wireless CO\textsubscript{2} Sensor}
    \cite{PASCOScientific2026} & \$288--300 & 1--2 hrs (leaf dies after excision) & Excised (stressed) & $\times$ & High \\
    \textbf{Vernier Go Direct CO\textsubscript{2}}
    \cite{VernierSoftware2026} & \$250--350 & 1--2 hrs (leaf dies after excision) & Excised (stressed) & $\times$ & High \\
    \midrule

    \multicolumn{6}{c}{\textbf{EMERGING LOW-COST SYSTEMS}} \\
    \midrule

    \textbf{Takaragawa et al. (2025)} 
    ~\cite{takaragawa2025establishment} & \$200 & single discrete measurements & Excised & $\times$ & Moderate \\
    \textbf{PhytoPulse (ESP32 + SCD40)} ~\cite{phytopulse2025} & \$50--100 & Hours--days & Excised & $\times$ & Moderate \\
    \midrule

    \multicolumn{6}{c}{\textbf{PhytoBits (THIS WORK)}} \\
    \midrule

    \textbf{PhytoBits} & \textbf{\$32--72} (See Table \ref{tab:component_cost_table}) & {Days--weeks (no labor; continuous autonomous monitoring)} & \textbf{Attached (living)} & \textbf{\checkmark} & \textbf{High} \\

    \bottomrule
  \end{tabularx}

  \footnotesize
  $^{\dagger}$ Distinguish C\textsubscript{3}/CAM:
    \checkmark = temporal pattern;
    $\sim$ = instantaneous or short period measurement;
    $\times$ = insufficient resolution.

  \caption{
    Comprehensive comparison of methods for observing and distinguishing C\textsubscript{3} vs. CAM photosynthesis. 
    Methods are organized by category (Research-Grade Instruments, Biochemical Assays, Educational Tools, Emerging Low-Cost Systems). 
    PhytoBits uniquely combines affordability, non-invasive attached-leaf measurement, multi-day monitoring capability, and 
    educational accessibility. Costs are approximate USD and may vary by region and configuration.
  }
\end{table}

\section{Photosynthesis theory and background}\label{section:photosynthesis_theory_and_background}
Photosynthesis occurs through continuous gas-exchange within the environment, different photosynthetic strategies produce distinct temporal patterns in these gas-exchange signals, among these, changes in \textbf{CO\textsubscript{2} concentration and humidity} in the leaf surroundings are what enable PhytoBits to distinguish between C\textsubscript{3} and CAM photosynthesis \cite{buschGuidePhotosyntheticGas2024}. 
This section provides the biological foundation needed to interpret the sensor data presented in subsequent sections.

\begin{figure}[H]
  \centering
  \includegraphics[width=0.9\linewidth]{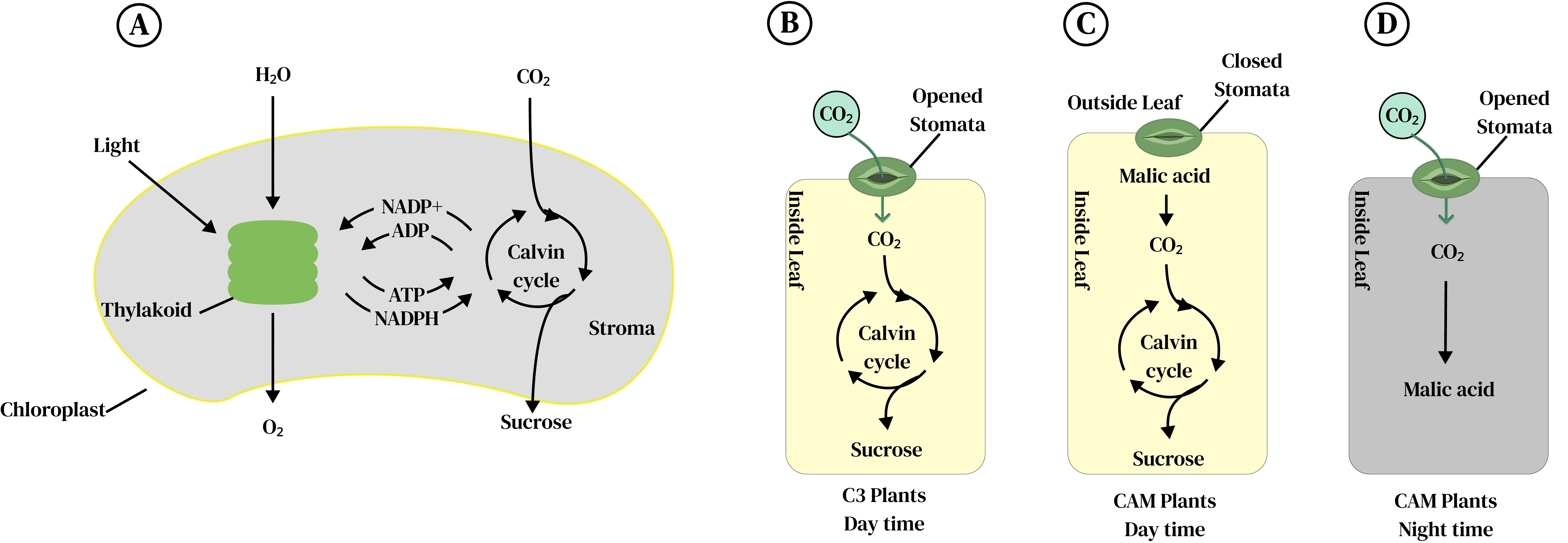}
  \caption{
  (A) light and dark reaction
  (B-D) CO$_2$ uptake strategies in C$_3$ and CAM plants across day-night cycles.
  }
  \Description{Different photosynthesis reactions and their pathways are described. The first part shows the light and dark reactions, whereas the rest of the parts show how CO$_2$ is absorbed in the plants with different strategies. C$_3$ and CAM plants have different mechanisms. Stomata opening and closing differs in these plants during day and night.}
  \label{fig:photosynthesis_reactions_pathways}
\end{figure}

\subsection{Overview of Photosynthetic Reactions}

\noindent\textbf{Light and carbon fixation fundamentals.} Photosynthesis occurs in two stages: light-dependent reactions (in thylakoids) produce ATP and NADPH as energy carriers while splitting water and releasing O\textsubscript{2}, and the light-independent Calvin cycle (in the stroma) uses ATP and NADPH as energy to fix CO\textsubscript{2} into sucrose. Every form of photosynthesis relies on light-dependent and Calvin-cycle reactions, but they vary in the \textit{timing and location} of carbon fixation \cite{Johnson2016Photosynthesis, buschGuidePhotosyntheticGas2024}.

\noindent\textbf{Stomatal control and detectable signals.} Plants acquire atmospheric \text{CO\textsubscript{2}} through stomata, microscopic pores primarily on the underside (abaxial surface) of leaves. When the stomata open, \text{CO\textsubscript{2}} is absorbed by leaf for photosynthesis, but water vapor simultaneously exits through transpiration, creating a fundamental trade-off. Guard cells regulate this trade-off by responding to light, \text{CO\textsubscript{2}} concentration, humidity, and other signals. When open, stomata produce two readily measurable signals to their surroundings: (1) \text{CO\textsubscript{2}} concentration decreases (due to leaf uptake), and (2) relative humidity increases (due to transpirational water release). The \textit{timing} of these signals differs predictably between photosynthetic pathways \cite{buschGuidePhotosyntheticGas2024}.

\subsection{Photosynthetic Pathways: Theory and Measurable Signals}
Plants use three major photosynthetic pathways, C\textsubscript{3}, C\textsubscript{4}, and CAM. They differ in how they balance carbon gain against water loss. The ancestral C\textsubscript{3} pathway dominates globally, especially in cooler or well-watered environments, whereas C\textsubscript{4} and CAM represent derived strategies that modify when and where \text{CO\textsubscript{2}} is fixed to improve performance in heat or drought.

\subsubsection{C\textsubscript{3} Photosynthesis: Diurnal CO\textsubscript{2} uptake}:
% \noindent\textbf{C\textsubscript{3} Photosynthesis (Prevalent but Water-Inefficient)}: 
C\textsubscript{3} photosynthesis dominates globally, most crops and houseplants, like tomatoes and pothos employ it. C\textsubscript{3} plants open their stomata during the day to permit rapid  uptake and close stomata during the night.

\subsubsection{CAM Photosynthesis: Nocturnal CO\textsubscript{2} uptake}
CAM evolved as an extreme water-conservation strategy \cite{winter2019ecophysiology, winter2022cam} for arid environments. CAM plants achieve 6-fold better water-use efficiency \cite{nobel1996high} than C\textsubscript{3} species through temporal separation: they open their stomata primarily at night (when cooler, it is less humid), fix \text{CO\textsubscript{2}} into organic acids (predominantly malic acid) that accumulate in vacuoles, then close their stomata during the day. During daytime, the stored malic acid releases \text{CO\textsubscript{2}} internally for use in the Calvin Cycle while water loss is minimized.

\subsubsection{Why Focus on C\textsubscript{3} vs. CAM, not C\textsubscript{4}?}
C\textsubscript{4} plants employ spatial separation (between mesophyll and bundle-sheath cells) rather than temporal separation, when monitored with gas-exchange sensors, C\textsubscript{4} plants exhibit diurnal CO\textsubscript{2} patterns identical to C\textsubscript{3} plants: daytime uptake and nighttime release. Therefore, \textbf{PhytoBits focuses on the C\textsubscript{3} and CAM distinction}, which is readily observable through gas-exchange timing and provides a pedagogically rich contrast \cite{keeley2003evolution, edwards2019evolutionary}.

\subsubsection{CAM Plasticity and Educational Value}

CAM physiology exists along a continuum rather than as a binary trait, creating valuable experimental opportunities:

\begin{figure}[H]
  \centering
  \includegraphics[width=0.7\linewidth]{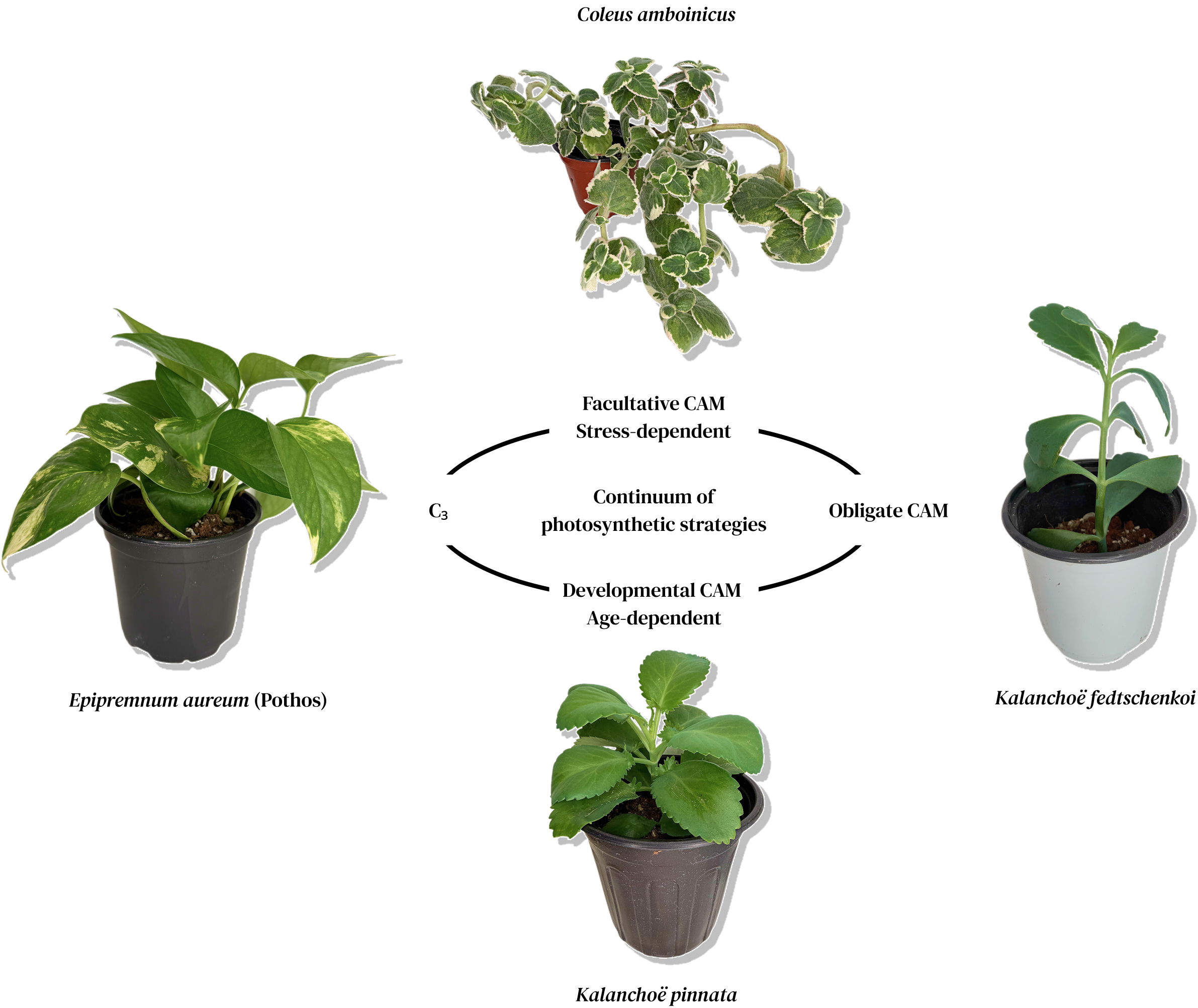}
  \caption{
  Representative plant species used for physiological comparison.
    (Left) \textit{E. aureum} (Pothos), C$_3$ plant with broad leaves.
    (Right) \textit{K. fedtschenkoi}, CAM succulent with bluish foliage.
    (Top) \textit{C. amboinicus}, Facultative CAM plant with soft and sometimes variegated leaves.
    (Bottom) \textit{K. pinnata}, Developmental CAM plant with thick, water-storing leaves.
  }
  \Description{Four photographs showing different plant species used in the study. Left side shows Epipremnum aureum (Pothos), a C3 plant with broad, heart-shaped green leaves. Right side shows Kalanchoe fedtschenkoi, a CAM succulent with small, bluish-gray fleshy leaves arranged along stems. Top image shows Coleus amboinicus, a facultative CAM plant with soft, rounded green leaves that may show variegation. Bottom image shows Kalanchoe pinnata, a developmental CAM plant with thick, oval-shaped succulent leaves that store water. The images illustrate the morphological diversity across different photosynthetic pathway types used for physiological comparison in the experiments.}
  \label{fig:Plants}
\end{figure}

\noindent\textbf{Constitutive (Obligate) CAM.}
Some plants, such as \textit{Kalancho\"e fedtschenkoi} (Lavender Scallops), rely exclusively on CAM photosynthesis throughout their entire life cycle, regardless of environmental conditions. 
This consistent metabolic character offers reliable baseline measurements that demonstrate the toolkit's ability to detect CAM patterns over weeks of continuous monitoring, and serve as positive controls. 
From an educational perspective, obligate CAM plants demonstrate that evolution can favor a single, highly specialized solution when environmental pressure is severe and consistent.

\noindent\textbf{Facultative CAM.}
Facultative CAM species, such as \textit{Coleus amboinicus}, exhibit reversible and graded shifts along the C\textsubscript{3}--CAM continuum in response to environmental stress \cite{winterFacultativeCrassulaceanAcid2014, winter2020constitutive, qiu2023mechanisms}. Under well-watered conditions, facultative CAM plants primarily rely on daytime \text{CO\textsubscript{2}} uptake characteristic of C\textsubscript{3} photosynthesis, while often retaining measurable nocturnal \text{CO\textsubscript{2}} fixation through low-level CAM activity \cite{winter2020constitutive}. When subjected to drought, salinity, or other stress applications, nocturnal CO\textsubscript{2} uptake becomes dominant, accompanied by reduced daytime stomatal conductance, reflecting an upregulation of CAM pathway components to conserve water. 
Upon stress relief, plants gradually return to C\textsubscript{3}-dominated gas-exchange patterns over several days, without permanent metabolic lock-in.
This plasticity reflects an adaptive strategy: facultative CAM plants maintain both metabolic pathways and activate CAM when stressed. 
From a classroom perspective, this reversibility is potentially transformative: students could design and conduct experiments by stressing the growing plants, monitoring gas-exchange dynamics to observe the gradual shift from C\textsubscript{3} to CAM over multiple days, then de-stressing and watching the reverse transition occur. 
This inquiry-based approach could directly connects plant physiology to student-driven environmental manipulation. Such experiments are impossible with research-grade instruments and invisible with traditional classroom methods.

\noindent\textbf{Developmental CAM.}
Certain succulent species, including \textit{Kalancho\"e pinnata}, undergo a metabolic transition as leaves mature and develop thicker succulent tissue, shifting from C\textsubscript{3} photosynthesis in young leaves to CAM in mature leaves\cite{wai2017temporal, abrahamTranscriptProteinMetabolite2016, winter2019ecophysiology}.
This developmental transition can be observed with PhytoBits by monitoring the young and mature leaves together (see Figure \ref{fig:developmental_cam}), %same leaf over days, 
offering students the rare opportunity to observe metabolic reprogramming.

\section{PhytoBits Toolkit Design}\label{section:PhytoBits_toolkit_design}
% All hardware designs, bill of materials, and firmware are available at \url{https://github.com/yourusername/PhytoBits}.

\begin{table}[H]
    \centering
    \caption{Toolkit Components and Associated Cost}
    \begin{tabular}{>{\raggedright\arraybackslash}p{4cm}lc}
    \hline
           &\textbf{Toolkit Component} & \textbf{Cost}\\\hline
    \multirow{5}{*}{Microcontrollers} &Arduino Uno R3 & \$ 27.60\\
           &Micro:bit Kit & \$ 19.22\\
   &Micro:bit Breakout Board &\$ 6.25\\
           &ESP32 & \$ 10.00\\
           &Arduino Mega 2560  & \$ 49.90\\\hline
   \multirow{2}{*}{Structural Frame Choice}&Malleable Aluminum Wire (30 cm long)&\$ 0.20\\
   &PLA Filament (1 kg Spool) &\$ 13.99\\\hline
   \multirow{2}{*}{Enclosure Material Choice}&1 Transparent Polyethylene Bag (12'' x 16'')&\$ 0.04\\
  &Parafilm Tape (3 cm wide)&\$ 0.10\\\hline
   Sensor&SCD41 CO\textsubscript{2} Sensor &\$ 21.62\\\hline
   Wiring&4 male to female jumper wires&\$ 0.23\\ \hline
   \multicolumn{2}{l}{\textbf{Total Cost Range}} & \textbf{\$ 32.09--72.09}\\\hline
    \end{tabular}
    
    \label{tab:component_cost_table}
\end{table}

The PhytoBits toolkit consists of a set of coordinated components, shown in Table \ref{tab:component_cost_table} with cost breakdown, designed for long-term and harmless monitoring of plant gas-exchange. 
The \textbf{leaf pod} is formed from a wire frame covered by a transparent flexible enclosure isolating a small region of the leaf with minimal physical disturbance. Inside the pod, \textbf{the sensing module} integrates CO\textsubscript{2}, humidity, and temperature sensors positioned to capture the microenvironment surrounding the leaf surface. A compact \textbf{microcontroller} collects and timestamps sensor readings and manages local data storage or transmission. Together with the \textbf{plant} under study, these elements form a self-contained sensing assembly capable of capturing extended gas-exchange dynamics in naturalistic conditions.

\begin{figure}[H]
  \centering
  \includegraphics[width=0.99\linewidth,trim=0 0 0 0, clip]{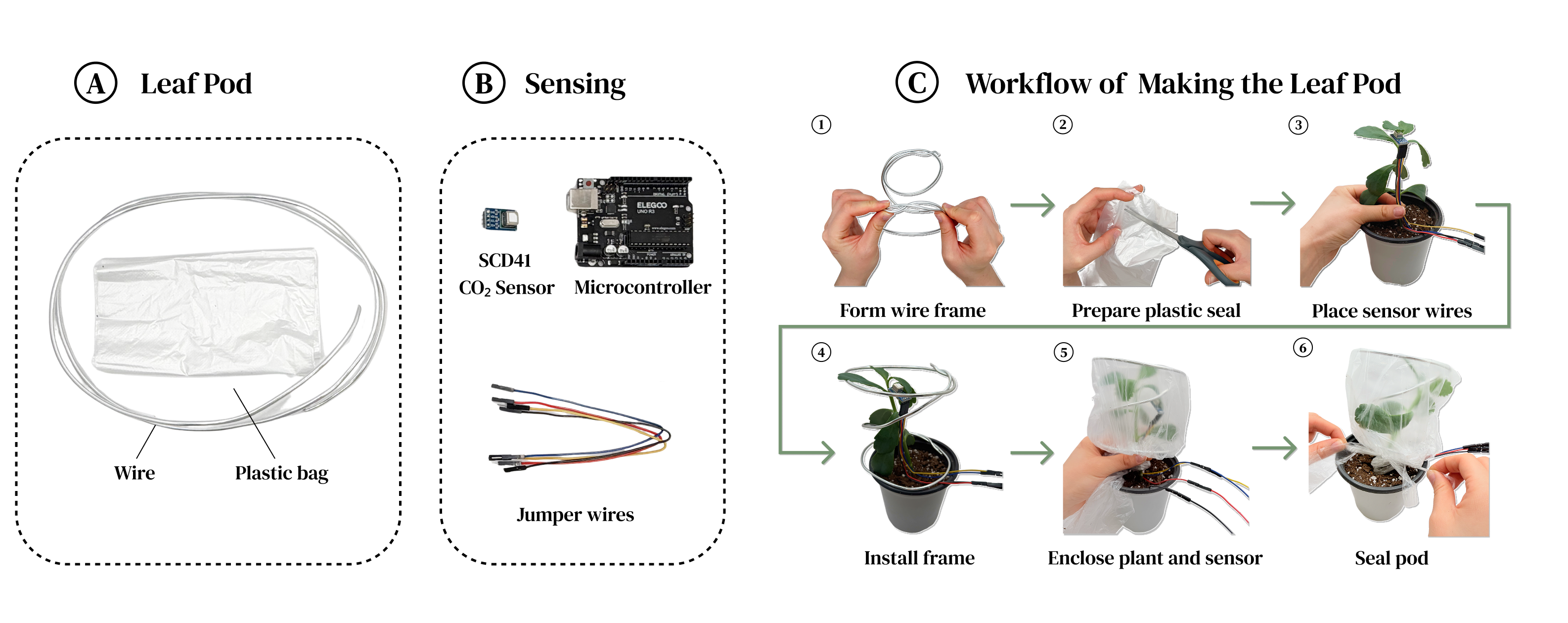}
  \caption{Leaf pod Components and Workflow for assembling leaf pod.
  (A) Components of the leaf pod enclosure, consisting of a flexible aluminum wire frame and a transparent plastic bag that forms a semi-sealed enclosure around the plant or a leaf (depending on the plant).
  (B) Sensing electronics, including an SCD41 CO\textsubscript{2} sensor, a microcontroller, and jumper wires for data acquisition and logging.
  (C) Assembly workflow: 
  1. Bend the wire into a spiral to form a support frame.
    2. Cut a strip (~1/2 inch) from the open end of the plastic bag to use for sealing later.
    3. Align the CO$_2$ sensor wires alongside the plant stem.
    4. Place the wire frame around the entire plant or a leaf (depending on the plant).
    5. Drape the plastic bag over the frame, enclosing the plant and leaf pod. Grasp the open end so only the stem and sensor wires remain outside.
    6. Wrap and tie the cut strip around the open end to seal the leaf pod.
  }
  \Description{Leaf pod components and assembly instructions: Panel A displays the physical components: a coiled aluminum wire frame for structural support and a clear plastic bag for enclosure. Panel B shows the electronic components: an SCD41 CO2 sensor module with a microcontroller and four jumper wires. Panel C presents a six-step assembly workflow illustrated with diagrams: Step 1 shows bending wire into a spiral frame; Step 2 shows cutting a thin strip from the bag opening for later sealing; Step 3 shows positioning sensor wires along the plant stem; Step 4 shows placing the wire frame around the plant or leaf; Step 5 shows draping the plastic bag over the frame to enclose the plant while keeping stem and wires outside; Step 6 shows wrapping and tying the cut strip around the opening to seal the chamber. The complete assembly creates a semi-sealed enclosure for measuring photosynthetic gas exchange.}
  \label{fig:leafpod_materials_and_assembly_workflow}
\end{figure}

\subsection{Micro-Environment Sensing Pod \label{subsec:leaf_pod}}
% figure for the leaf pod without sensors
Traditional photosynthesis measurement approaches face limitations in cost, invasiveness, and temporal resolution. To enable accessible, non-invasive, long-term monitoring of leaf-level gas-exchange, we developed the micro-environment sensing pod as a core component of the PhytoBits toolkit. The pod creates a semi-sealed enclosure around individual leaves to enable continuous tracking of CO\textsubscript{2} and humidity as indicators of photosynthetic activity.

% One of the core physical components of the PhytoBits toolkit is the lightweight, micro-environment sensing pod. It is designed to create a sealed enclosure around individual leaves or a group of leaves. 
The key highlights of the pod are its utilization of easily accessible materials, ease of assembly without special equipment, and adaptability to diverse plant morphologies.
The structural frame of the pod is constructed from malleable aluminum wire or PLA filament material commonly used in 3D printing. Both materials enable the creation of a cylindrical helical form, as shown in Figure \ref{fig:leafpod_materials_and_assembly_workflow}. This pod provides sufficient structural integrity while remaining lightweight enough to avoid stressing the enclosed plant. Material choice can be tailored to the experimental context: steel wire offers greater sturdiness and adjustable height, making it suitable for classroom demonstrations and for leaves positioned closer to the ground, whereas PLA is lighter and more flexible, enabling unobtrusive placement around upper canopy leaves without requiring additional external support.

We first create the cylindrical frame by bending the wire. The cylindrical frame is wrapped with a transparent polyethylene bag to form the leaf pod. The transparency of the polyethylene maintains the light exposure to enclosed leaves, preserving photosynthetic activity of the plants while creating the gas-exchange boundary necessary for micro-environment sensing. We tie the bag with our cut out strip of bag as shown in Figure \ref{fig:leafpod_materials_and_assembly_workflow}. In some cases, parafilm can be used to seal the pod\ref{subsubsec:sealing methods}. The sensor wires and the stem are aligned so they can be sealed together. This modular, low-cost pod design enables deployment across multiple plants simultaneously and can be rapidly reconfigured as experimental needs evolve. These minimal fabrication requirements support the toolkit's goal of democratizing access to plant data and phenotyping capabilities for resource-constrained research settings, educational settings, and community science applications.

To support long-duration sensing in naturalistic settings, we avoid pumps and chamber mixing fans typically used to compute instantaneous steady-state gas-exchange rates. Instead, we adopt a continuous CO\textsubscript{2} monitoring strategy. This reduces mechanical complexity, lowers deployment overhead, and allows us to capture extended temporal patterns in plant gas-exchange without disturbing the plant or its environment.

\begin{figure}[H]
  \centering
  \includegraphics[width=0.45\linewidth]{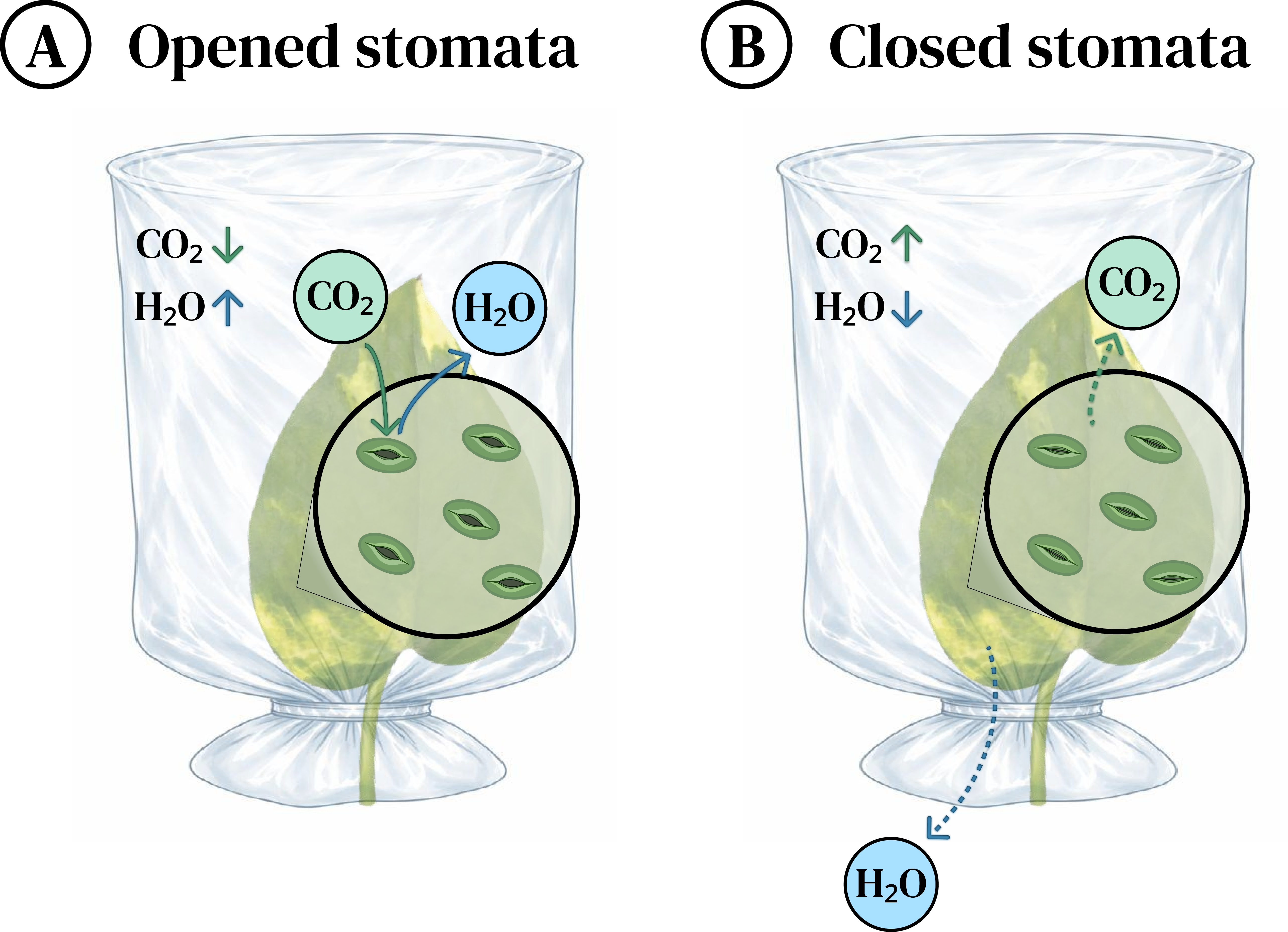}
  \caption{
  (A) When stomata are open,  is absorbed by the leaf and water vapor is released, leading to a decrease in CO₂ concentration and an increase in humidity within the leaf pod.
  (B) When stomata are mostly closed, CO₂ accumulates due to respiration and limited diffusion, while humidity reduced because of the water vapor going outside the leaf pod.
  }
  \Description{detailed process diagram showing the physiological processes and chemical exchanges that happen when the stomata open and close inside the leaf pod. Panel A shows how CO\textsubscript{2} is absorbed and water vapor is released through the stomata when it is open, and when it is closed, the CO\textsubscript{2} concentration rises due to respiration and humidity decreases due to water vapor going out of the pod. }
  \label{fig:toolkit_working_principle}
\end{figure}

This microenvironment enables observation of plant physiological signals. As shown in Figure \ref{fig:toolkit_working_principle}, when the plant opens its stomata, the CO\textsubscript{2} from the pod is absorbed by the stomata and the water vapor from the plant is released into the pod from the stomata increasing the humidity and decreasing the CO\textsubscript{2} concentration. Similarly, when the stomata is closed, the humidity goes out from the pod and the CO\textsubscript{2} increases due to plant respiration, thus inverting the pattern of open stomata. Even though the stomata is closed, the respiration still happens since it is independent of stomata closure. Furthermore, there is evidence that gas diffusion still happens when stomata is closed shown by stomatal conductance  \cite{nighttime_stomamtal_conductance}.  For C\textsubscript{3} plants, the stomata open during daytime, while it closes during the night. Conversely, for CAM plants, the stomata open during the night and closes during the daytime. 

% #######################################################################################

% \noindent\textbf{Observable signal:} in a leaf enclosure, C\textsubscript{3} plants produce a characteristic diurnal pattern:
% \textbf{Daytime:} Stomata open:  decreases humidity rises, \textbf{Nighttime:} Stomata close:  accumulates, humidity decreased

% \noindent\textbf{Observable signal:} In a leaf enclosure, CAM plants exhibit an inverted pattern:
% \textbf{Nighttime:} Stomata open:  decreases, humidity rises, \textbf{Daytime:} Stomata closed:  increases, humidity decreased

% Figure \ref{fig:principle} illustrates how stomatal behavior shapes the gas-exchange signals detected in the leaf enclosure. 

% #########################################################################################

\subsection{Design Parameters and Considerations}
\subsubsection{Leaf Pod Sealing Methods\label{subsubsec:sealing methods}}
Two practical approaches for sealing the micro-environment sensing pod around the leaf and sensor wiring: a bag-tying technique and a parafilm wrap were evaulated. In the bag-tying method, the transparent enclosure is formed from a polyethylene bag and the stem and sensor wires are routed through a narrow neck at the pod opening. A strip of the same polyethylene material is used to form a constrictive knot around the stem and wires, producing a simple mechanical seal. In the parafilm method, the same initial bag configuration is used, but the bag neck is wrapped and sealed with parafilm rather than with a tied plastic strip. Both procedures permit rapid assembly with low-cost materials and require minimal tooling.

Comparative experiments showed differences in sealing performance and susceptibility to external perturbation. Parafilm provided a tighter seal, as evidenced by control experiments in which room ambient fluctuations produced little to no influence on the pod microenvironment. The tying method produced an effective seal under most conditions, but in poorly ventilated spaces with sustained human occupancy, the pod records were sometimes perturbed by changes in background CO\textsubscript{2}. This ventilation and occupancy sensitivity is described in detail in Section \ref{sec:human_activty}. These observations indicate a trade-off between convenience and airtightness practitioners should consider when choosing a method.

For experiments requiring maximal isolation from ambient variability, parafilm sealing is preferred owing to its superior airtightness. For classroom demonstrations, rapid field deployments, or situations where ease of assembly and reusability are prioritized, the tying method remains an acceptable and robust option provided background conditions are monitored.

\vspace{-0.15in}
\subsubsection{Pod Volume and Its Impact on Measurements}
\begin{figure}[H]
  \centering
  \includegraphics[width=0.9\linewidth]{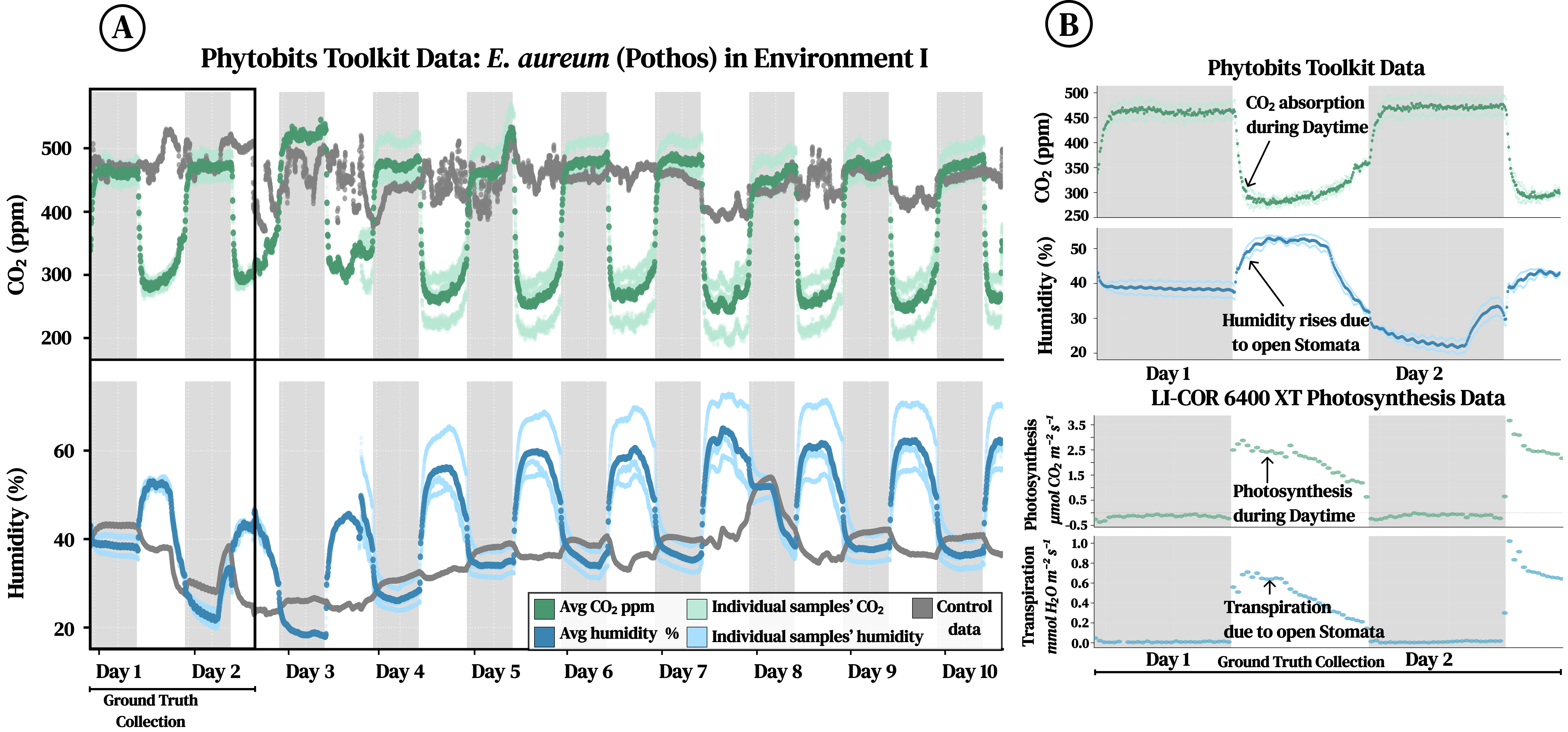}
  \caption{(A) PhytoBits' capture of C\textsubscript{3} photosynthesis in \textit{E. aureum}. CO\textsubscript{2} depletion occurs during daytime with concurrent humidity increase. (B) Comparison of PhytoBits and LI-COR 6400 XT measurements. The inverse relationship between LI-COR photosynthetic activity and PhytoBits CO\textsubscript{2} concentration validates the result.}

  \Description{plot showing CO\textsubscript{2} and humidity patterns of E. aureum captured by Phytobits, showing C\textsubscript{3} pattern (CO\textsubscript{2}) absorbed during day with simultaneus increase in humidity and reverse pattern during the night. The data matches with the photosynthesis trend shown by LI-COR 6400 XT which validates the PhytoBits results.}
  \label{fig:C3}
\end{figure}

We tested four species: \textit{K. fedtschenkoi}, \textit{E. aureum}, \textit{C. amboinicus}, and \textit{K. pinnata}. Pod selection was driven by leaf size, fragility, and the strength of photosynthetic signals. \textit{E. aureum} produced strong signals from a single large leaf and was measured in the same 600–800 cm$^{3}$ pod using one leaf. These results can be observed in Figure \ref{fig:C3}. For \textit{K. fedtschenkoi}, individual leaves are small and produce weak signals, so the entire plant was enclosed in a leaf pod (cylindrical volume of 600–800 cm$^{3}$). The results can be observed in Figure \ref{fig:Obligate_CAM}. For \textit{C. amboinicus} we placed a branch containing many small leaves into the 600–800 cm$^{3}$ pod to obtain sufficient leaf area. These results can be observed in Figure \ref{fig:facultative_CAM}. \textit{K. pinnata} required two approaches: a fragile, very small young leaf was measured inside a small bag (no metal frame; volume: 100 cm$^{3}$), whereas a mature leaf was accommodated in a larger sized bag (volume: 350 cm$^3$). This is due to the leaves being relatively small and fragile (the young leaf area was 8 cm$^2$ while the mature leaf area was 40 cm$^2$). The results for \textit{K. pinnata} can be observed in Figure \ref{fig:developmental_cam}.

These observations indicate that a 600–800 cm$^{3}$ cylindrical pod is appropriate for most measurements when adequate leaf area is available. When individual leaves are too small or signals are weak, use multiple leaves or enclose the whole plant; for very small or fragile material, use a smaller, frame-free bag to avoid mechanical damage.

% \subsubsection{Trade-offs Between Usability and Measurement Quality}

\subsection{Sensor Selection and Calibration} 
The toolkit integrates a SCD41 CO\textsubscript{2} concentration sensor \cite{sensirion2023scd4x}. It is a high-accuracy sensor breakout board that uses photoacoustic Non-Dispersive Infrared (NDIR) technology to achieve a ``true'' CO\textsubscript{2} measurement. It is coupled with a built in humidity and temperature sensor for on-chip compensation for accurate readings, and its smaller size allows for a compact form factor of the toolkit. To ensure comparability between toolkit readings and ground truth measurements, we disabled the CO\textsubscript{2} sensor's internal self calibration feature. Our experiments require free registration of ambient concentrations below 400 ppm, and disabling automatic self calibration prevents the sensor from forcing readings toward an assumed baseline, preserving low-end sensitivity. Second, the sensors with an updated lower measurement range of 0 ppm were calibrated in an ambient outside environment. This procedure ensured the sensors were able to log values below 400 when the concentration went below 400. 

An EZO O\textsubscript{2} sensor\cite{atlasScientificEZO_O2_2023} was evaluated during initial prototyping but was excluded from the final toolkit for two reasons. First, oxygen concentration provided minimal additional information for determining photosynthetic timing, as O\textsubscript{2} is produced during the light-dependent reactions and therefore primarily reflects daytime activity. Second, incorporating the O\textsubscript{2} sensor substantially increased both system cost and mechanical complexity.

\subsection{Cross Platform Microcontroller Compatibility}
The democratization of scientific tools through low-cost, open-source hardware has transformed both formal education and community science ~\cite{pearce2012building,oellermann2022open}. Microcontroller-based platforms have proven particularly effective in enabling hands-on experimentation across diverse user communities ~\cite{chan2021low}. However, different microcontrollers vary systematically in timing accuracy, data storage capacity, connectivity options, and software ecosystems, which can limit usability for specific audiences. To ensure that PhytoBits remains accessible and reliable across educational levels and research contexts, we conducted a comprehensive cross-platform evaluation using three widely adopted microcontroller families, each addressing distinct user needs:

\begin{itemize}
\item \textbf{ESP32}: Ideal for researchers, advanced students and long-term field studies, it is often selected as the primary platform for scientific deployments due to its built-in Wi-Fi and Bluetooth connectivity, high-resolution timing capabilities, robust data-logging performance, and sufficient memory for extended autonomous operation. Majority of the experiments in Section \ref{main_section_results} were done using ESP32.
\item \textbf{Arduino UNO and MEGA}: Arduino offers a mature software ecosystem and straightforward programming model. Its ubiquity in makerspaces, and community science initiatives makes it the natural choice for educators implementing PhytoBits in informal learning environments \cite{chan2021low,camprodon2019smart}.

For data collection from both ESP32 and Arduino, we used a script for logging the serial data and uploaded it periodically to the cloud (we used AWS for data storage).
\item \textbf{Micro:bit}: Designed for and used extensively in K-12 classrooms globally, the Micro:bit provides a visual programming interface (MakeCode) alongside traditional text-based coding. While resource-constrained compared to ESP32 and Arduino, its age-appropriate design and built-in sensors make it a promising entry point for younger students (ages 8-14) \cite{sentance2017teaching}. The results from Figure \ref{fig: other_species} were collected through Micro:bit and the results from Figure \ref{fig:light_conditions} were collected through Micro:bit on-chip storage.
\end{itemize}

This cross-platform approach enables instructors and researchers to select platforms matched to their technical expertise, budget constraints, and pedagogical goals.
It also validates consistent photosynthetic signatures across platforms, which demonstrates that underlying biological phenomena are robust to instrumental variation, increasing confidence in crowd-sourced data from heterogeneous sensor networks ~\cite{bonney2009citizen,chan2021low}.
By standardizing sensor protocols and data pipelines for each microcontroller family, PhytoBits maintains scientific rigor while lowering barriers to entry. This design philosophy 
%aligns with broader movements toward open science hardware and inclusive STEM education,%
ensures plant physiological research is not limited to well-resourced laboratories but can be conducted by students, educators, and community scientists worldwide ~\cite{oellermann2022open,pearce2012building}.

\section{Measurement Feasibility and Validation Study}

\subsection{Ground Truth Measurements\label{subsubsec:ground_truth_measurements}}

Ground-truth measurements of photosynthetic activity were obtained using two complementary, established techniques.

First, gas-exchange measurements were collected using a LI-6400 Portable Photosynthesis System (LI-COR, Inc., Lincoln, NE, USA). The CO$_2$ supply was provided by a compressed CO$_2$ tank equipped with an air regulator set to an output pressure of 200 PSIG. The Infrared Gas Analyzer (IRGA) was fitted with a clear chamber head to allow ambient light to pass through during measurements. For each plant measured with the LI-COR, we recorded standard outputs including net CO\textsubscript{2} assimilation rate and transpiration. Photosynthetic rate (Photo) and transpiration rate (Trmmol) were measured at a reference CO\textsubscript{2} concentration of 400~$\mu$mol~mol$^{-1}$ and an air flow rate of 500~$\mu$mol~s$^{-1}$. Desiccant and soda lime were replaced every 6-20 hours, depending on researcher availability, to maintain stable humidity and CO$_2$ scrubbing performance. These measurements were taken at the leaf level and served as the primary quantitative reference for photosynthetic fluxes during both daytime and nighttime sampling windows.

Second, for species known to employ CAM, we quantified malic acid concentration in leaf tissue using acid titration at paired daytime and nighttime sampling points. Leaves were sampled 2~hours before lights-on (dawn) and 2~hours before lights-off (dusk) \cite{hu2024transcriptomic, winter2020constitutive}. 
Immediately after excision, each leaf was flash-frozen in liquid nitrogen to halt metabolic activity. To accomplish this, liquid nitrogen was poured into a beaker, a piece of aluminum foil was floated on the surface, and the leaf was placed on the foil until fully frozen. Frozen samples were transferred to cryogenic tubes, placed on dry ice, and stored at $-$20~\textdegree C until analysis.

For titration, a subsample of frozen tissue was removed, weighed, and ground thoroughly using a mortar and pestle. The homogenate was transferred to a beaker and mixed with 10~mL of distilled water. The initial pH was measured using an Apera Instruments PH60S spear-tip pH meter. Diluted NaOH solutions (0.0005-0.005 mol L$^{-1}$; concentration selected based on expected acidity) were then added incrementally while mixing, and pH was recorded until the solution reached pH~6.5 \cite{winter2020constitutive}. The volume of NaOH required was then used to calculate malic acid content as percent by mass using equation \eqref{eq:malic}, assuming 2 mols of NaOH neutralizes 1 mol of malic acid, where $V_{\mathrm{NaOH}}$ is the volume of NaOH in liters (read from burette), $C_{\mathrm{NaOH}}$ is the concentration of NaOH in mol~L$^{-1}$ (value retrieved via dilution), $M_{\mathrm{malic}}$ is the molar mass of malic acid $(134.09$~g~mol$^{-1})$ \cite{malic_acid}, and $m_{\mathrm{leaf}}$ is the mass of the leaf sample in grams (weighed on a scale).

\begin{equation}
    \%\,\text{malic acid}
    = \left( \frac{V_{\mathrm{NaOH}} \cdot C_{\mathrm{NaOH}} \cdot M_{\mathrm{malic}}}{2 \cdot m_{\mathrm{leaf}}} \right) \times 100
\label{eq:malic}
\end{equation}
\noindent    

All ground-truth measurements were time-stamped and synchronized with the PhytoBits sensor logs to enable direct comparison. Together, these ground-truth techniques validated the reliability and interpretability of the PhytoBits measurements.

% \subsubsection{PH Meter?}

\begin{figure}[H]
    \centering
    \includegraphics[width=0.7\linewidth]{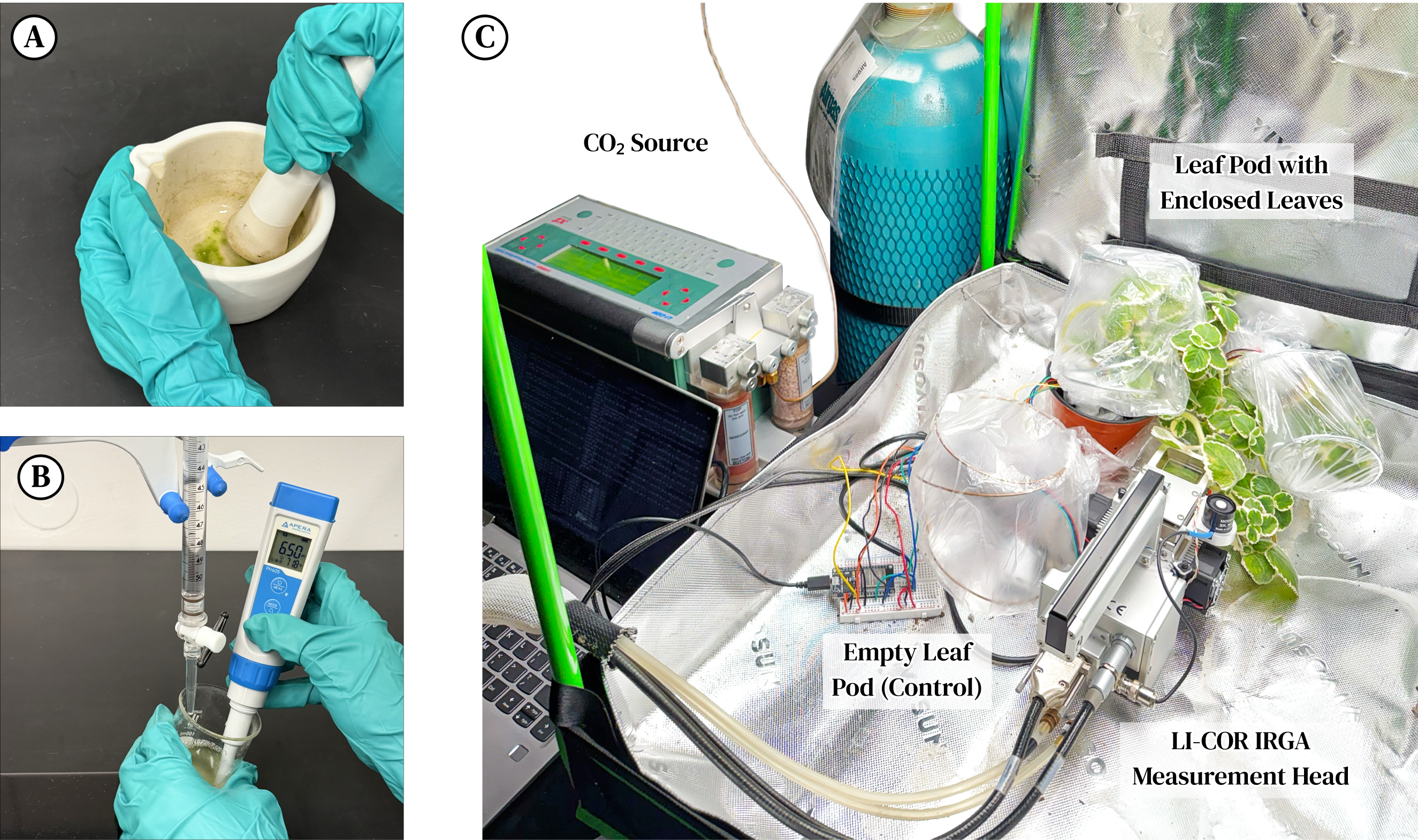}
    \caption{
        Ground-truth measurements for validating PhytoBits-based CO$_2$ sensing.
        (A) Leaf tissue is ground using a mortar and pestle to extract organic acids.
        (B) The extract is titrated with NaOH until reaching pH 6.5 to quantify titratable acidity.
        (C) Experimental setup showing the PhytoBits sensing system operating alongside a LI-COR 6400 XT gas-exchange system for ground-truth comparison, including the CO$_2$ cylinder and pressure regulator, artificial growth light, custom leaf pod with enclosed leaves (experiment) and an empty leaf pod (control), the LI-COR IRGA measurement head, and the data acquisition laptop.
    }
    \Description{Three-panel figure showing ground-truth validation methods. Panel A shows a plant leaf tissue being ground in a white mortar and pestle to extract organic acids for titratable acidity measurement. Panel B shows a pH meter reading 6.5 next to a beaker containing leaf extract being titrated with sodium hydroxide solution to quantify acid content. Panel C shows the complete experimental setup with labeled components: a CO2 gas cylinder with pressure regulator for controlled atmosphere, an artificial growth light providing illumination, two custom leaf pods (one enclosing plant leaves for experimental measurements and one empty as control), a LI-COR 6400 XT gas exchange system with its infrared gas analyzer measurement head clamped to leaves, the PhytoBits sensing system with microcontroller and CO2 sensor, and a laptop computer for data acquisition from both systems simultaneously for validation comparison.}
    \label{fig:ground_truth_collection}
    % \label{fig:titration}
  % \hfill
\end{figure}
% \todo[]{description}

%deleted DHT
% The {DHT22 (Temperature and Humidity) sensor} is a widely used, low-cost digital sensor for measuring relative humidity and temperature. It uses a capacitive sensor for humidity and an $\text{NTC}$ thermistor for temperature. The sensor offers a high resolution of $0.1$ and good accuracy over its operating range ($-40^\circ \text{C}$ to $80^\circ \text{C}$ and $0$ to $100\%\text{ RH}$). It also allows for a compact form factor due to its size and can be easily integrated with different microcontrollers.

\subsection{Controlled Environment Setup} \label{sec:env_setup}

All experiments for the \textit{E. aureum} (Pothos) and \textit{K. fedtschenkoi} were conducted indoors in three distinct controlled environments to evaluate toolkit performance. The three environments were chosen to capture realistic variation a deployable toolkit might encounter and test whether the CO\textsubscript{2}, humidity, and temperature signals remain diagnostic of photosynthetic timing across contexts. The experiments with other plants were also done in at least one of these environments.

\textbf{Environment 1} %(D112) 
was a small room where the growth chamber for the plant experiments was placed. The growth chamber equipped with grow lights had a peak intensity of 1100 lux inside the leaf pod, and around 2400 lux inside the growth chamber but outside the leaf pod. Inside the leaf pod, temperatures ranged from 21-22~\textdegree C during the dark period and increased to 26-28~\textdegree C when the lights were on; these patterns were consistent across both control pods and pods containing leaves. Temperature also exhibited a characteristic waveform pattern, with approximately 0.5~\textdegree C fluctuations during the daytime due to fluctuations introduced by the building ventilation system. 
This environment represented a lower volume, higher plant to air ratio condition. All of the LI-COR ground truth measurements were taken in this environment. 

\textbf{Environment 2} %(L477)
was a larger growth chamber in a meeting room, fitted with grow lights capable of housing a substantial number of plants. This chamber provided a high plant density condition and larger air volume, which tests the toolkit's sensitivity to distributed CO\textsubscript{2} uptake. The peak light intensities inside and outside the leaf pod in this environment were 820 lux and 1250 lux. Inside and outside the leaf pod, temperatures ranged from 18-20~\textdegree C during the dark period and increased to 22-25~\textdegree C when the lights were on; these patterns were consistent across both control pods and pods containing leaves.

\textbf{Environment 3} %(L470)
was an intermediate sized enclosure located in a third room with artificial lighting. The peak light intensities inside and outside the leaf pod in this environment were around 850 lux and 1600 lux. Inside the leaf pod, temperatures ranged from 22-24~\textdegree C during the dark period and increased to 24-26~\textdegree C when the lights were on; these patterns were consistent across both control pods and pods containing leaves. 

For each environment, we maintained consistent, documented control variables like light intensity and temperature (slightly dependent on time of day and real world temperature). Furthermore, as shown in Figure \ref{fig:ground_truth_collection}, we had an empty leaf pod in each of the three environments as our control reading. This control was placed near the plants to ensure robustness. This multi-environment setup allowed the evaluation of the reliability of the PhytoBits toolkit under different conditions and its ability to distinguish the photosynthesis phenotypes under different conditions.

\section{Feasibility of observing photosynthesis phenotypes}\label{main_section_results}
In this section, we present the Phytobits data from different plants capturing different photosynthetic patterns. 

\begin{figure}[!h]
  \centering
  \includegraphics[width=0.99\linewidth]{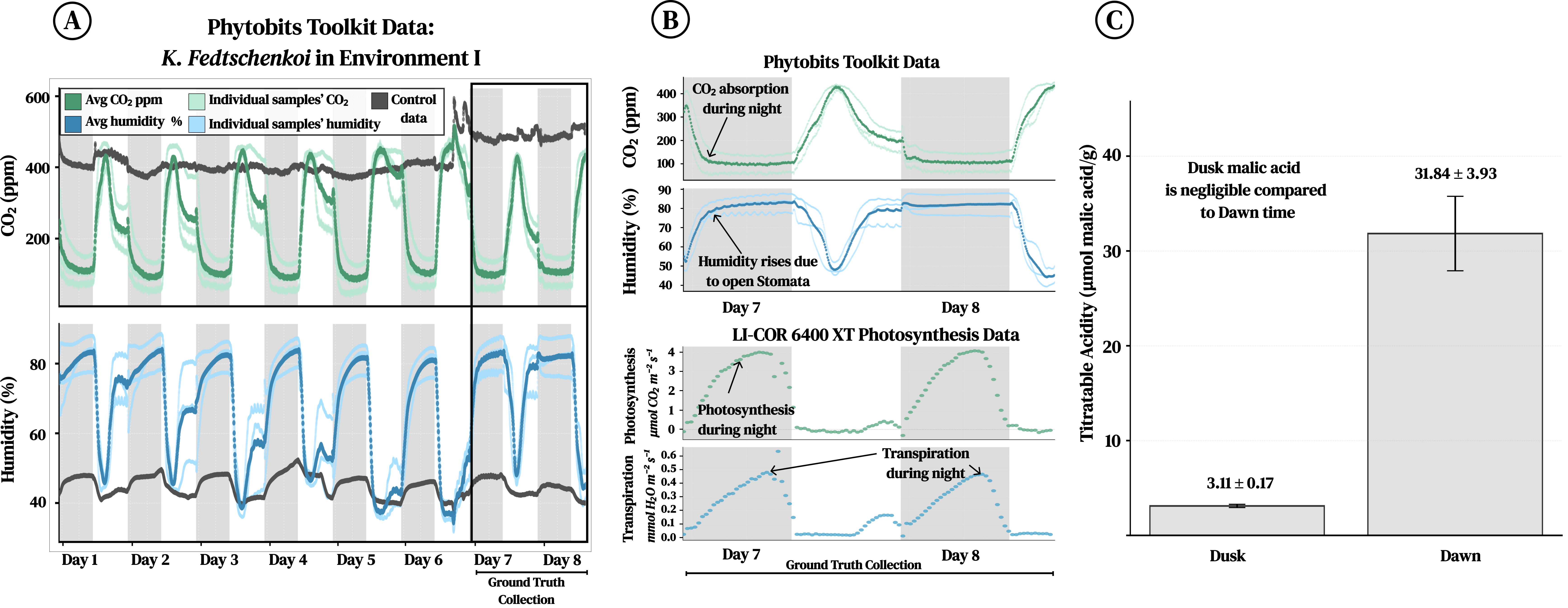}
  \caption{(A) PhytoBits' capturing CAM photosynthesis in \textit{K. fedtschenkoi}. CO\textsubscript{2} depletion occurs during night with concurrent humidity increase. (B) Comparison of PhytoBits and LI-COR 6400 XT measurements. The inverse relationship between LICOR photosynthetic activity and PhytoBits CO\textsubscript{2} concentration validates the result. (C) Comparison of leaf Malic acid levels during dawn and dusk to further validate the toolkit results.}
  \Description{PhytoBits capturing an Obligate CAM signal from a \textig{K. fedtschenkoi} plant. 
   CO\textsubscript{2} concentrations decrease and humidity increases during the night, 
  consistent with daytime stomatal opening and Obligate CAM photosynthesis.}
  \label{fig:Obligate_CAM}
\end{figure}

\subsection{Observation of C\textsubscript{3}} 
\label{subsec:C3_and_cam_results}

The diurnal gas-exchange dynamics in \textit{E. aureum} (Pothos) was monitored over an 10-day period using PhytoBits. The time series exhibited a characteristic C\textsubscript{3} photosynthetic pattern. The CO\textsubscript{2} concentrations within the pod decreased during the light period while the relative humidity increased, whereas the opposite trend was observed during the dark period (elevated CO\textsubscript{2} and reduced humidity). These diel shifts reflect daytime stomatal opening, CO\textsubscript{2} assimilation, and increased transpiration, followed by nighttime stomatal closure. Note that we faced some data loss in two of our Pothos plants on day three while adding new plants in the controlled environment. This happened due to accidental touch to some wires which got disconnected. Thus, the individual samples are not clearly visible on the plot for that time duration, however, it was fixed before Day 4.

To validate these observations,  ground-truth measurements using a LI-COR gas-exchange system were collected from Day 1 to Day 2, with operational details provided in Section \ref{subsubsec:ground_truth_measurements}. The LI-COR data aligned closely with the PhytoBits measurements, confirming photosynthetic CO\textsubscript{2} uptake occurred primarily during the light period and transpiration rates were substantially higher during the day. These results demonstrate PhytoBits reliably captures canonical C\textsubscript{3} physiological behavior, including daytime stomatal conductance and water-vapor release.

\subsection{Observation of Obligate CAM} \label{obligate_cam_results}
To observe PhytoBits' ability to detect obligate CAM photosynthesis, we monitored \textit{K. fedtschenkoi}, a constitutive CAM species, under identical environmental conditions. In contrast to the C\textsubscript{3} pattern observed in \textit{E. aureum} (Pothos), \textit{K. fedtschenkoi} exhibited an inverted gas-exchange profile characteristic of obligate CAM metabolism (Figure \ref{fig:Obligate_CAM}A). CO\textsubscript{2} concentrations within the pod decreased during the dark period while relative humidity increased, indicating nocturnal stomatal opening and CO\textsubscript{2} fixation. During the light period, CO\textsubscript{2} levels rose and humidity remained relatively stable, reflecting daytime stomatal closure, a water-conserving strategy that distinguishing CAM from C\textsubscript{3} photosynthesis.

PhytoBits measurements were comparatively validated with ground-truth data collected using a LI-COR 6400 XT gas-exchange system and leaf malic acid quantification via acid titration (Section \ref{subsubsec:ground_truth_measurements}). The LI-COR data confirmed nocturnal photosynthetic activity, showing an inverse relationship between nighttime CO\textsubscript{2} assimilation rates and PhytoBits CO\textsubscript{2} concentrations (Figure \ref{fig:Obligate_CAM}B). Additionally, malic acid levels, a key biochemical marker of CAM, were significantly elevated in leaf samples collected at the end of the dark period compared to those harvested during the day (Figure \ref{fig:Obligate_CAM}C), consistent with nocturnal CO\textsubscript{2} fixation and organic acid accumulation in the vacuole.
These converging lines of evidence demonstrate PhytoBits accurately resolves the temporal inversion of gas-exchange that defines obligate CAM photosynthesis.

\subsection{Observation of Facultative CAM} \label{facultative_CAM_section}
\begin{figure}[!h]
  \centering
  \includegraphics[width=\linewidth]{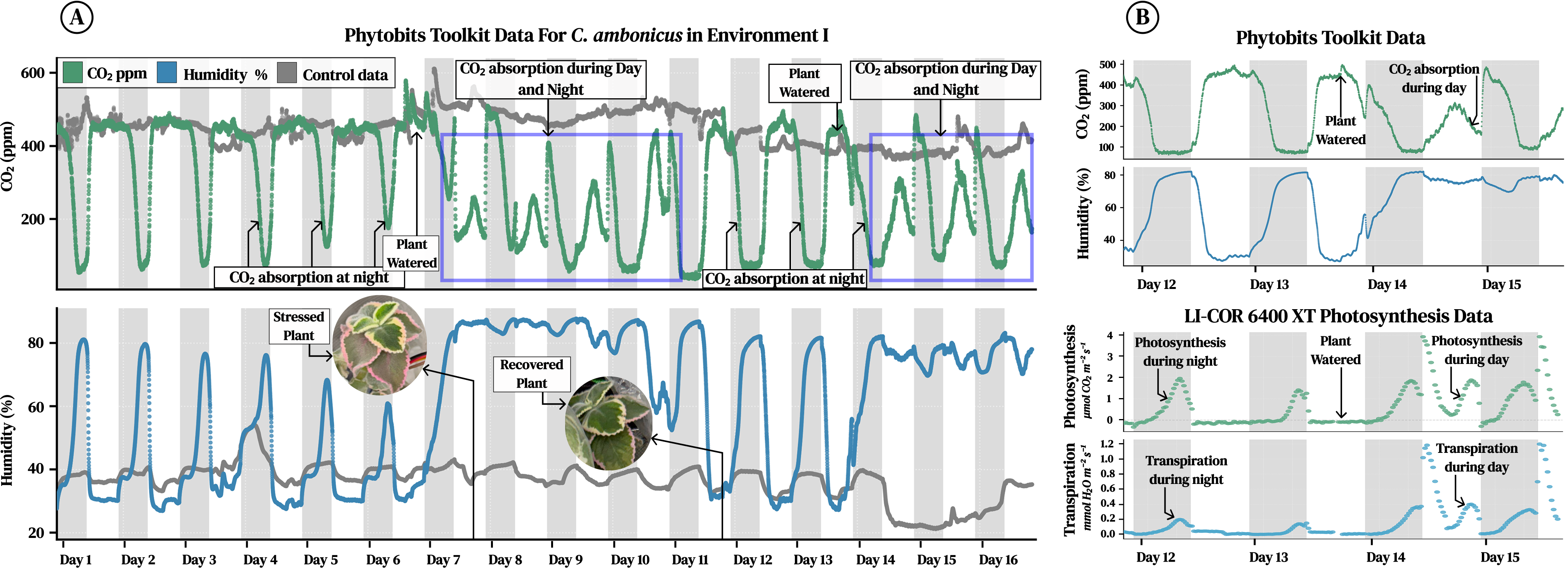}
  \caption{(A) PhytoBits' capture of Facultative CAM. \textit{C. amboinicus} exhibits CAM photosynthetic patterns in the first week, followed by a mixture of CAM and C\textsubscript{3} pattern upon watering. (B) Comparison of PhytoBits and LI-COR 6400 XT measurements. The inverse relationship between LI-COR photosynthetic activity and PhytoBits CO\textsubscript{2} concentration validates the result.}
  \Description{PhytoBits capturing an Facultative CAM signal from a \textig{C. amboinicus} plant. The plant initially shows CAM data, and upon watering the plant, it shows a mixture of  C\textsubscript{3} and CAM photosynthesis patterns. After a few days, it shifts to CAM again. Comparison with LI-COR 6400 XT data validates the Phytobits results. }
  \label{fig:facultative_CAM}
\end{figure}

To demonstrate PhytoBits' ability to detect facultative CAM behavior, we observed \textit{C. amboinicus} under different water availability conditions. Facultative CAM species exhibit metabolic flexibility, switching between C\textsubscript{3} and CAM photosynthesis or exhibiting both the patterns together depending on environmental stress. During the first few days of the plot, the plant was not watered for over two weeks, inducing drought stress. Thus, a clear CAM pattern is visible from Days 1 to 6. We could also observe the leaves of the plant showed stress indicators such as color change. \textit{C. amboinicus} is known to exhibit this response to drought and light stress \cite{stress_produces_anthocyanin_ref, anthocyanin_gives_red_pigment}. The plant was watered on Day 6, and under well-watered conditions (Day 7 - Day 10), the plant displayed a mixture C\textsubscript{3} and CAM pattern: CO\textsubscript{2} concentrations within the pod decreased during the light period as well as during the night with concurrent increases in relative humidity, indicating daytime stomatal opening and photosynthetic CO\textsubscript{2} assimilation (Figure \ref{fig:facultative_CAM}A). We also observed that the pinkness of the leaves reduced after watering the plant, showing signs of recovery.

Measuring the Plant photosynthesis through LI-COR 6400 further validates the mixed CAM and C\textsubscript{3} pattern detected by PhytoBits. Upon watering, daytime  CO\textsubscript{2} uptake increased along with the existing nighttime CO\textsubscript{2} uptake (Figure \ref{fig:facultative_CAM}B). The temporal dynamics captured by PhytoBits align with established biochemical markers of CAM engagement reported in the literature ~\cite{abrahamTranscriptProteinMetabolite2016,mingPineappleGenomeEvolution2015}.
These results suggest PhytoBits can reliably distinguish between constitutive C\textsubscript{3} photosynthesis and stress-induced CAM in facultative species.

\begin{figure}[!h]
\centering
\includegraphics[width=\linewidth]{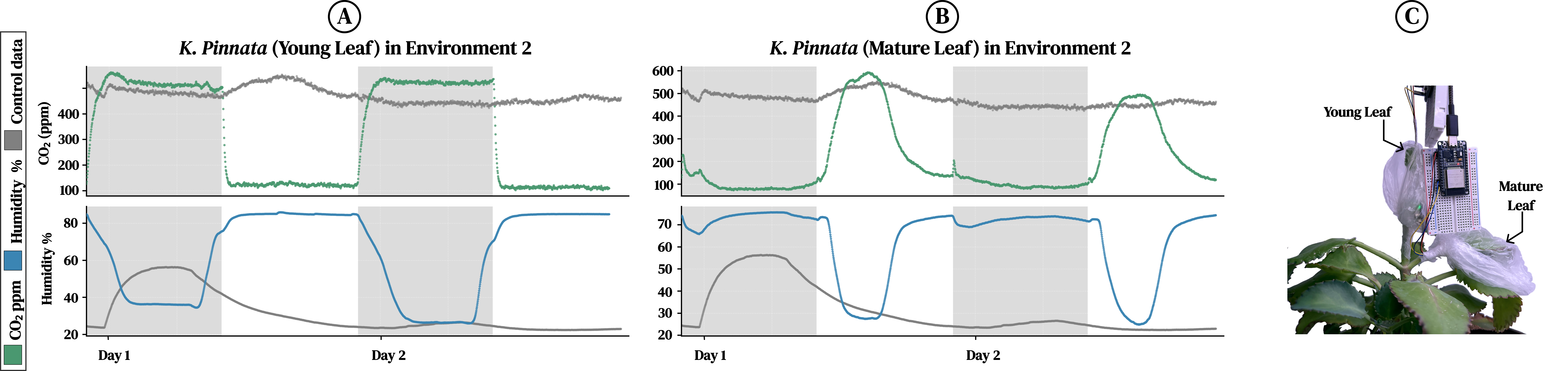}
\caption{(A) PhytoBits' capture of C\textsubscript{3} photosynthesis in a young leaf of \textit{K. pinnata}.  CO\textsubscript{2} depletion occurs during daytime with concurrent humidity increase. (B) PhytoBits' capture of CAM photosynthesis in a mature leaf of the same plant during the same time. (C) Locations of young and mature leaves in the plant.}
\Description{Three-panel figure demonstrating developmental CAM photosynthesis in \textit{K. pinnata}. Panel A show CO2 concentration and humidity over in a young leaf exhibiting C3 photosynthesis, with CO2 levels decreasing during daytime hours while humidity increases, indicating stomatal opening and active photosynthesis during the day. Panel B shows a corresponding graph for a mature leaf of the same plant exhibiting CAM photosynthesis, with a different pattern where CO2 changes occur primarily at night when stomata open, followed by daytime CO2 depletion from stored organic acids. Panel C shows a photograph of the K. pinnata plant with arrows or labels indicating the positions of the young leaf (displaying C3 behavior) and mature leaf (displaying CAM behavior) measured in panels A and B, illustrating how photosynthetic pathway shifts with leaf development on the same individual plant.}
\label{fig:developmental_cam}
\end{figure}

\subsection{Observation of Developmental CAM}\label{developmental_cam_section}

% text for developmental cam
To further demonstrate PhytoBits' ability to capture different photosynthetic patterns in different plant species, we test it on \textit{K. pinnata}, a plant known to exhibit developmental CAM. The toolkit was instrumented on the top most young leaf and mature leaf from the same plant. Figure \ref{fig:developmental_cam} shows photosynthetic patterns of both the young leaf (A) and the mature leaf (B). Both leaves from the same plant exhibit completely reverse patterns. The young leaf exhibits C\textsubscript{3} behavior with  CO\textsubscript{2} absorption during day time, whereas the mature leaf shows CO\textsubscript{2} absorption at night. (Figure \ref{fig:developmental_cam}C) shows the setup for capturing developmental CAM. We did not use the metal frame of the PhytoBits toolkit and used parafilm to seal the leaf stem and sensor inside the bag since the leaves were too small and fragile. However, the photosynthetic patterns were still captured. 

These results (captured with the help of the ESP32 Microcontroller) demonstrate clear discrimination among photosynthetic phenotypes. However, the data doesn't just depend on the physiological conditions, there are a lot of environmental factors such as water levels, light schedules, and human activity, that can affect the data. As a result, environmental variable modulating PhytoBits' measurement must be taken into account.

\subsection{Environmental Manipulations and Their Effects} \label{env_manipulations}

Building on the validation presented in \ref{subsec:C3_and_cam_results}, we examine how controlled manipulations of the growth environment influence the physiological signals captured by PhytoBits and how these perturbations inform data interpretation and experimental design. With emphasis on the advantages of continuous, high temporal resolution sensing, this subsection is organized into targeted analyses of human activity and room ventilation, watering regimes and drought stress, alterations to light period timing, a jet-lag style light shift demonstration, and comparisons between natural and artificial illumination. For each manipulation we describe the experimental protocol, summarize the salient changes observed in the PhytoBits time-series data, and discuss implications for robust phenotype detection and for deploying the toolkit in different environments. Together, these discussions provide a systematic account of environmental factors that affect the data, and highlight different types of experiments that could be done with the PhytoBits toolkit.

\subsubsection{Human Activity and Room Ventilation Effects}\label{sec:human_activty}
\begin{figure}[!h]
  \centering
  \includegraphics[width=\linewidth]{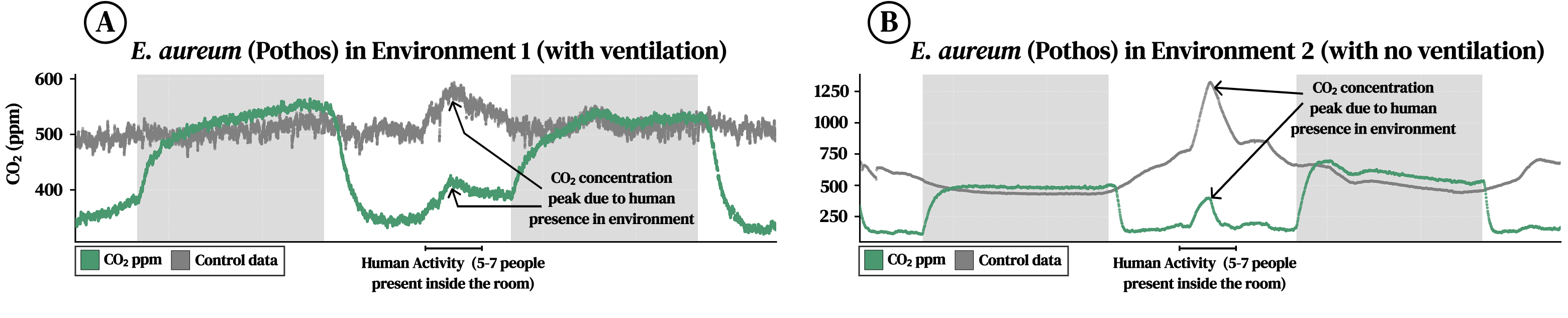}
  \caption{(A) Effect of human presence on CO\textsubscript{2}  data in a ventilated room. (B) Effect of human presence on CO\textsubscript{2} data in a non-ventilated room.}
  \Description{Plots showing human activity can interfere with the data if the room is not ventilated. Panel A shows how occupancy of humans may affect the data even when the room is ventilated, similarly, Panel B shows how CO2 data can rise very quickly when human occupancy is present in a room that is not ventilated.}
  \label{fig:effect_of_human_activity}
\end{figure}

Figure \ref{fig:effect_of_human_activity} illustrates how transient changes in room occupancy and ventilation regime produce pronounced distortions in the CO\textsubscript{2} gas-exchange data recorded by PhytoBits. In Environment 1, a well ventilated room, when the human occupancy exceeded five for a long duration, the data starts getting affected. Both the control, and the plant CO\textsubscript{2} show spikes during the peak occupancy timings. However, through the plot, one can still determine if the CO\textsubscript{2} absorption is happening during the day or night. In Environment 2, we observed that closing the room door (the only means of ventilation) and preventing the prolonged presence of five or more people led to a sustained rise in ambient CO\textsubscript{2} propagating into the pod-level recordings. These occupancy-induced CO\textsubscript{2} excursions manifest as apparent increases in plant CO\textsubscript{2} uptake and can obfuscate the real patterns when not accounted for. By contrast, when the room door was opened to restore proper room ventilation, the ambient CO\textsubscript{2} concentration remained near steady state and the artifact was substantially reduced (similar to Environment 1), demonstrating simple ventilation changes eliminates the confounding signal.

Since PhytoBits provides continuous, long term measurements, these occupancy and ventilation artifacts are detectable and temporally localizable, which permits both post hoc correction and proactive experimental control. In practice we recommend several measures to mitigate such effects: instrumenting the room with at least one ambient CO\textsubscript{2} sensor to record background fluctuations; annotating periods of human presence in the data log; and maintaining controlled ventilation. Implementing these controls improves the reliability of inferred plant physiological responses.

\begin{figure}[!h]
  \centering
  \includegraphics[width=0.99\linewidth]{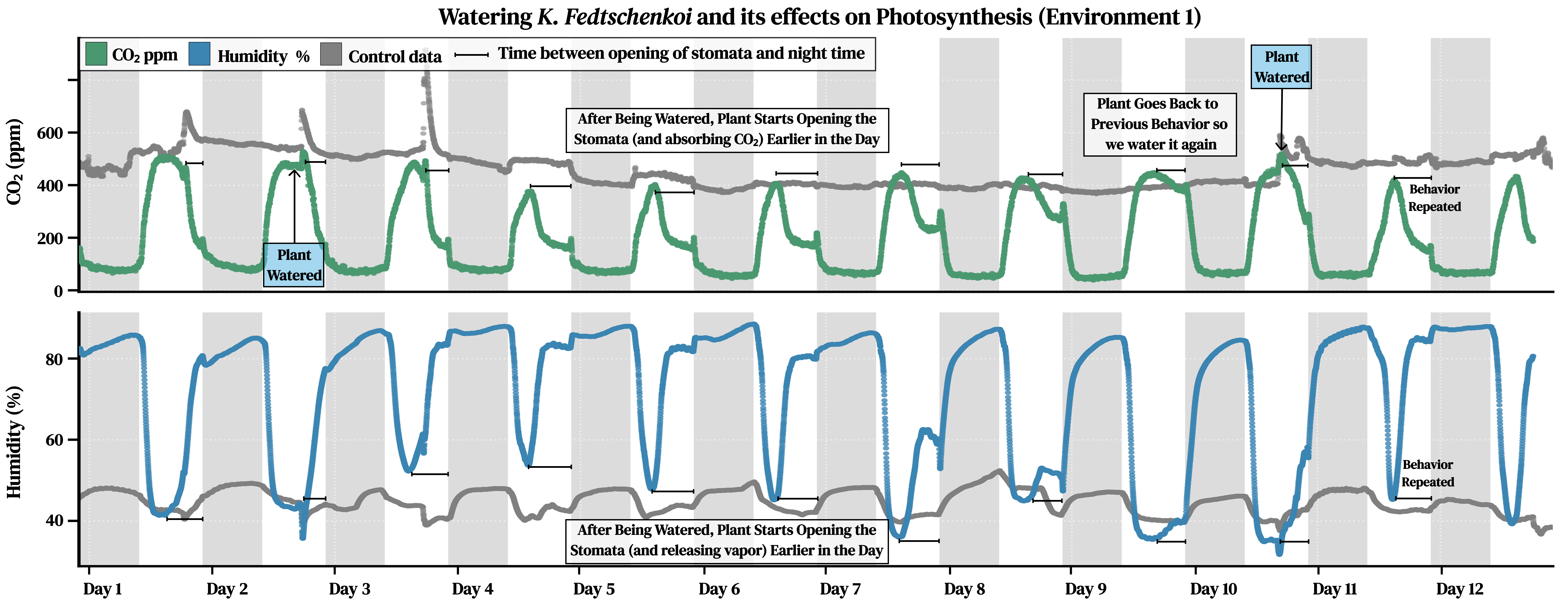}
  \caption{Prior to watering the \textit{K. fedtschenkoi}, the plant exhibits a CAM-dominated pattern characterized by nocturnal CO\textsubscript{2} uptake and nighttime stomatal opening. Following watering, CO\textsubscript{2} absorption begins earlier in the photoperiod, indicating a slight shift toward daytime stomatal opening and increased photosynthetic confidence under improved water availability.}
  \Description{Plot showing CO2 concentration and humidity measurements in \textit{K. fedtschenkoi} over multiple days before and after watering. Before watering, CO2 uptake occurs primarily at night (CAM pattern). After watering (marked by vertical line), CO2 absorption shifts to begin earlier in the evening, indicating increased daytime stomatal activity with improved water availability.}
  \label{fig:watering}
\end{figure}

\begin{figure}[!h]
\centering
\includegraphics[width=0.75\linewidth]{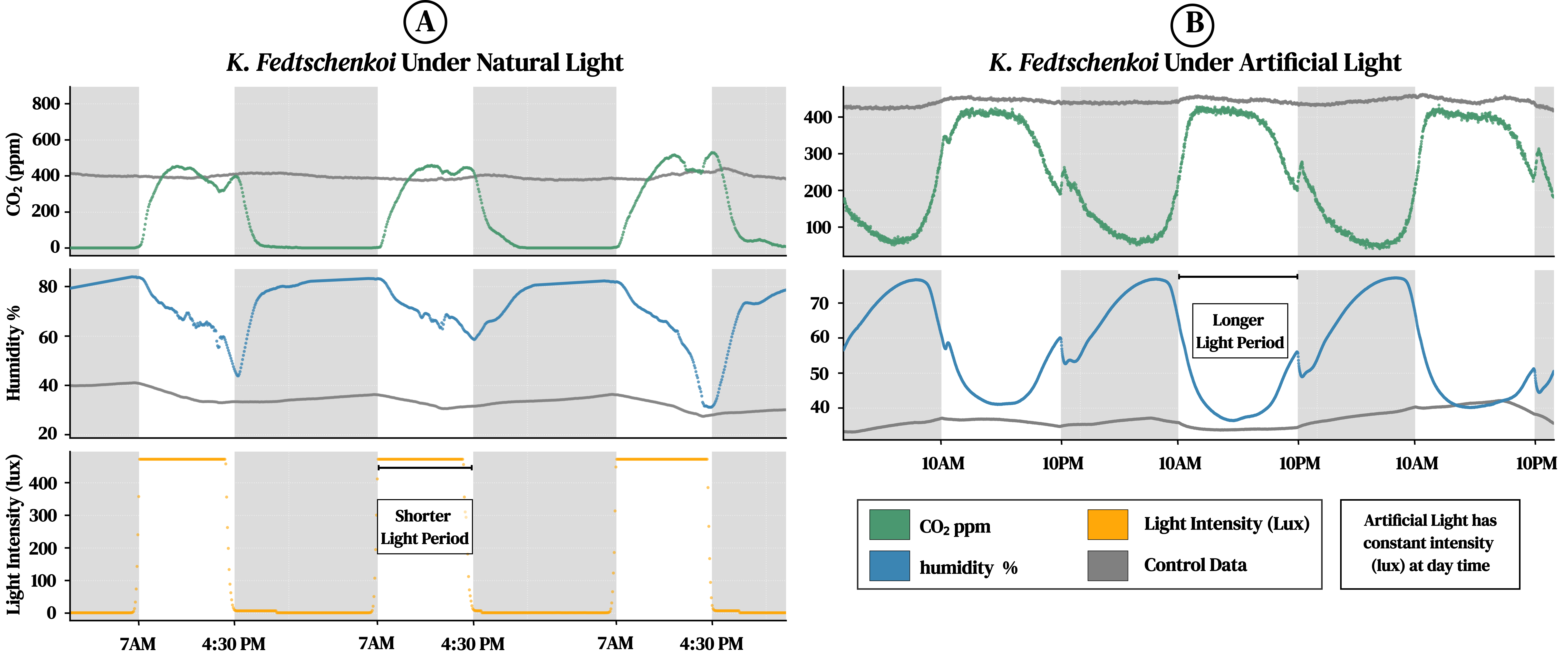}
\caption{Gas-exchange responses of \textit{K. fedtschenkoi} under (A) Natural Light, and (B) Artificial Light, captured with PhytoBits.}
\Description{Two plots comparing CO2 and humidity patterns in K. fedtschenkoi under natural light (panel A) versus artificial light (panel B), showing differences and similarities in photosynthetic gas exchange responses between lighting conditions. Phytobits is able to capture CAM behavior in both environmental conditions.}
\label{fig:light_conditions}
\end{figure}

\subsubsection{Watering Regimes and drought stress}\label{watering_regime_subsection}
% D112 Kalancho\"e, more confident to open the stomata after being watered
Figure \ref{fig:watering} shows the response of \textit{K. fedtschenkoi} to watering events recorded with PhytoBits. Under drought conditions the plant exhibited a CAM-like profile with nocturnal CO\textsubscript{2} uptake; following watering, CO\textsubscript{2} absorption clearly began a bit earlier in the day, consistent with increased daytime stomatal opening. This gradual advance in the onset of daytime CO\textsubscript{2} uptake highlights the sensitivity of PhytoBits to subtle, environmentally driven transitions and demonstrates the platform's ability to resolve rapid physiological responses to changes in water availability. 

We have also observed another physiological response to watering captured by PhytoBits in Section \ref{facultative_CAM_section} where watering the plant led to induction of C\textsubscript{3} photosynthetic pattern. Similarly, many other experiments can be explored to see how different plants react to watering events. 

\subsubsection{Light Period Manipulation and Reversal}
% \textbf{Jet-Lag Demonstration}

\begin{figure}[H]
\centering
\includegraphics[width=0.99\linewidth]{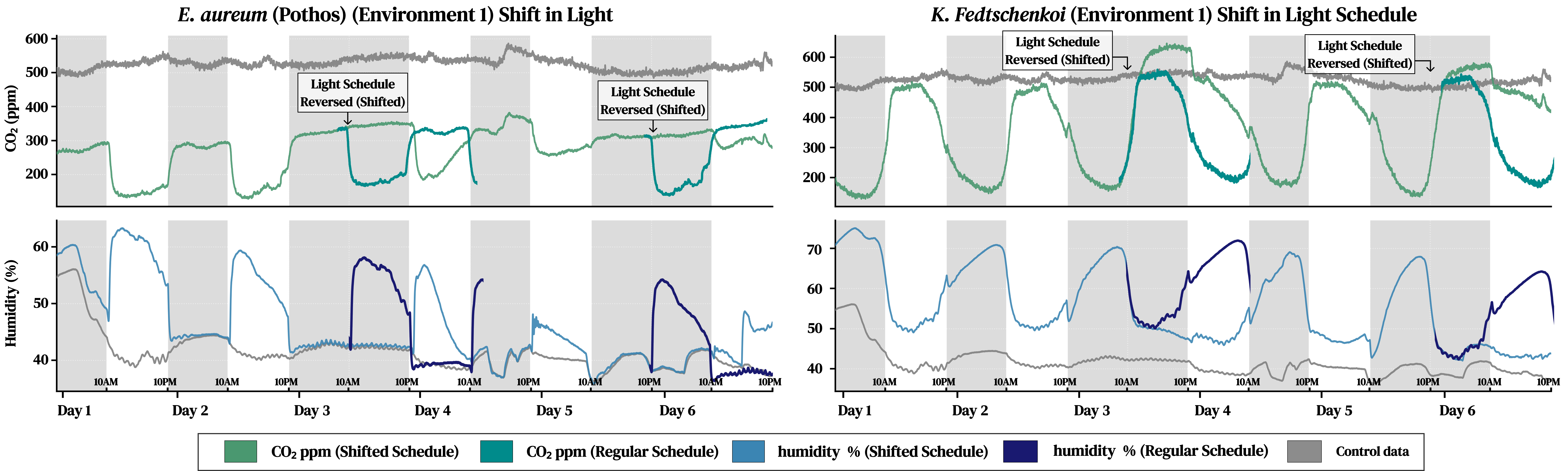}
\caption{Gas-exchange responses of \textit{E. aureum} (Pothos) and \textit{K. fedtschenkoi} under an inverted light schedule, illustrating plant dynamics like delayed reactions captured with PhytoBits.}
\Description{PhytoBits sensor result with light schedule being manipulated. It shows how E. aureum adapts quickly to light but reduces its day time CO2 uptake gradually. On the other hand, K. fedtschenkoi takes some time to adapt to the new light schedule evidenced by its delayed reaction. }
\label{fig:jetlag}
\end{figure}

In Environment 1 we performed a controlled light-period inversion to probe circadian entrainment dynamics under a 10:00-22:00 photoperiod (day) and 22:00-10:00 scotoperiod (night). The experiment was conducted on two co-located specimens of \textit{E. aureum} (Pothos) and \textit{K. fedtschenkoi} with Arduino UNO microcontrollers and the light schedule was reversed at the annotated time in Figure (dark shading signifies light off) \ref{fig:jetlag}.

The \textit{E. aureum} (Pothos) responded immediately to each photoperiod inversion: following the first reversal it retained its canonical C\textsubscript{3} behavior (absence of nocturnal CO\textsubscript{2} uptake and daytime uptake resumed), and upon the second reversal it again tracked the imposed light cues without an obvious phase lag. Notably, the magnitude of daytime CO\textsubscript{2} uptake in \textit{E. aureum} (Pothos) diminished gradually over the course of the experiment, and daytime uptake following each reversal was lower than the pre-inversion baseline. This progressive reduction in daytime assimilation suggests an attenuation of photosynthetic capacity or altered stomatal/conductance dynamics over time (e.g., as a consequence of cumulative stress or changing of light periods), a nuance that sparse sampling would have missed.

By contrast, \textit{K. fedtschenkoi} exhibited a transient, jet-lag-like response to each light reversal. Immediately after the first inversion, the \textit{K. fedtschenkoi} failed to exhibit its normal nocturnal CO\textsubscript{2} fixation; instead, CO\textsubscript{2} uptake appeared during the subsequent photoperiod, a clear transient mismatch between external light cues and internal CAM timing. Within ~24 hours, the plant reestablished the characteristic CAM profile (nocturnal fixation aligned to the new dark interval), but on repeating the reversal the same ~1-day delay recurred. Thus, \textit{K. fedtschenkoi} shows pronounced entrainment inertia relative to \textit{E. aureum}: it requires one or more cycles to realign nocturnal stomatal timing to the altered photoperiod.

 PhytoBits’ continuous, high-temporal-resolution records permitted direct visualization and quantification of both the phase shifts and the changes in uptake magnitude shown in Figure \ref{fig:jetlag}, enabling estimation of entrainment time constants and detection of subtle, short-lived deviations that would be difficult to capture with discrete LI-COR sampling alone. These results underscore the value of continuous monitoring for dissecting the interaction between endogenous timing and environmental forcing.

\subsubsection{Natural vs. Artificial Light Conditions}
Figure \ref{fig:light_conditions} compares thePhytoBits time series collected from an obligate CAM specimen \textit{K. fedtschenkoi}, under two illumination regimes (Natural light vs. Artificial light). PhytoBits resolves canonical CAM behavior in both conditions, nocturnal CO\textsubscript{2} uptake and daytime stomatal closure are clearly apparent, but the temporal shape and magnitude of the signal differ with the illumination regime. Under the controlled, artificial light schedule the light irradiance is constant across the pre-set time, thus, the daytime interval appears broader and any residual daytime CO\textsubscript{2} dynamics are temporally extended; by contrast, under natural daylight the effective daytime window is shorter because light levels fall earlier in the evening, producing a compressed/earlier-terminating daytime signature and modest shifts in the timing and amplitude of both daytime and nocturnal CO\textsubscript{2} dynamics. These differences likely reflect the direct effects of irradiance timing on stomatal behavior and other processes that influence nocturnal fixation. Importantly, despite these modulations, the principal CAM phenotype is robustly detected in both regimes, and continuous PhytoBits monitoring enables quantitative comparison of how specific illumination schedules map onto CAM gas-exchange patterns. 

\subsection{Robustness Across Environments}
\begin{figure}[!h]
  \centering
  \includegraphics[width=0.99\linewidth]{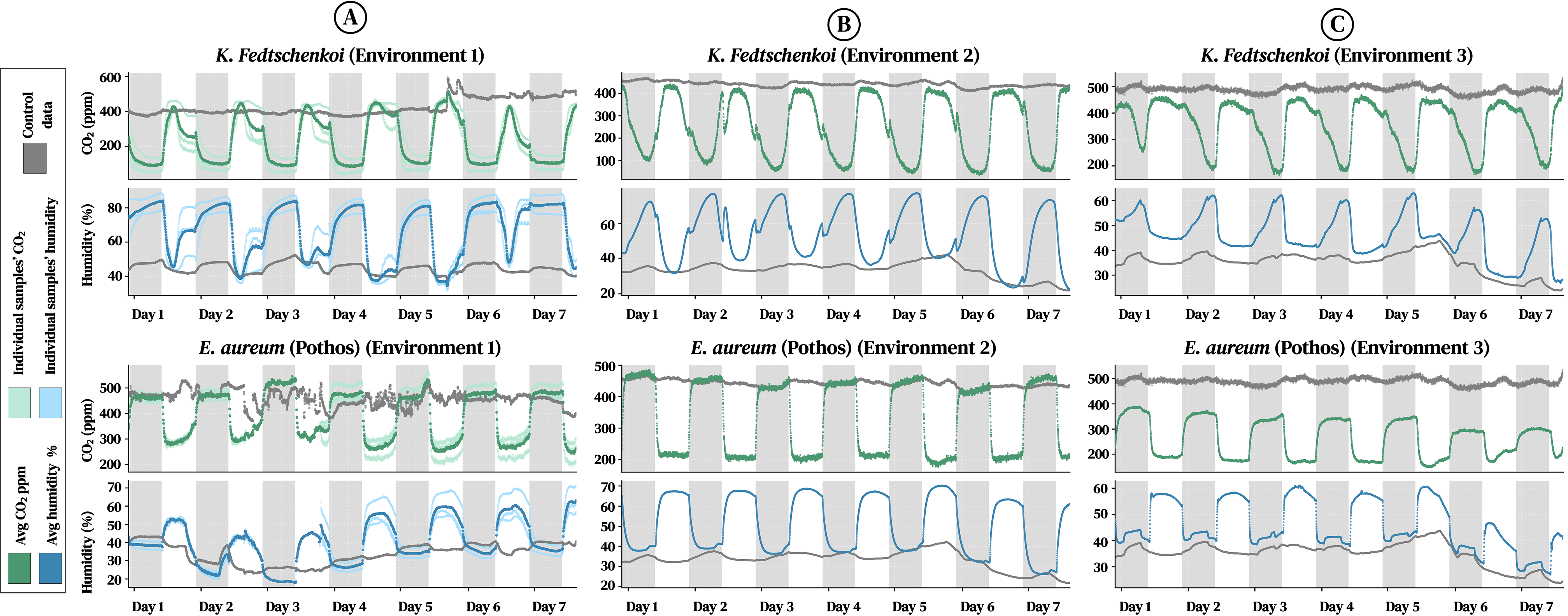}
  \caption{Robustness of PhytoBits across environments. 
The same plant species (\textit{K. fedtschenkoi} and \textit{E. aureum}) measured in three distinct environments exhibits consistent C$_3$/CAM signatures, indicating that PhytoBits produces stable physiological patterns despite environmental variation.}
  \Description{Multi-panel plot showing CO2 and humidity measurements for two plant species across three different environments. The top row shows K. fedtschenkoi exhibiting consistent CAM photosynthesis patterns (nighttime CO2 uptake) in all three locations. The bottom row shows E. aureum displaying consistent C3 photosynthesis patterns (daytime CO2 depletion) across the same three environments. Despite environmental differences, each species maintains its characteristic photosynthetic signature, demonstrating the PhytoBits' ability to capture stable physiological patterns across varied conditions.}
  \label{fig:env}
\end{figure}

Figure \ref{fig:env} summarizes a cross-environment validation in which representative C\textsubscript{3} and CAM specimens were monitored with PhytoBits in the three environments specified in Section \ref{sec:env_setup}. Although environmental parameters such as air volume, background ventilation, irradiance profile, and ambient humidity differed between locations, the core inherent temporal signatures that distinguish C\textsubscript{3} from CAM physiology were preserved. Specifically, C\textsubscript{3} plants consistently showed daytime CO\textsubscript{2} depletion while CAM plants exhibited nocturnal CO\textsubscript{2} uptake; the principal differences across environments were limited to modest changes in the amplitude and sharpness of CO\textsubscript{2} and humidity peaks.

These results indicate that PhytoBits delivers stable phenotype classification across heterogeneous deployment contexts while retaining sensitivity to environment-driven modulation of signal magnitude and temporal form. In practice this means that PhytoBits can reliably identify photosynthetic strategy even when absolute uptake rates vary because of local microclimate or instrumentation placement.  

% For experimentalists and field deployers we recommend reporting environmental metadata and, when possible, colocating an ambient sensor for background  and humidity. % Doing so facilitates normalization and comparison across sites while preserving the system's ability to detect biologically meaningful shifts in timing and intensity of gas-exchange.

\subsection{PhytoBits Results From Other Plant Species}

\begin{figure}[H]
  \centering
  \includegraphics[width=\linewidth]{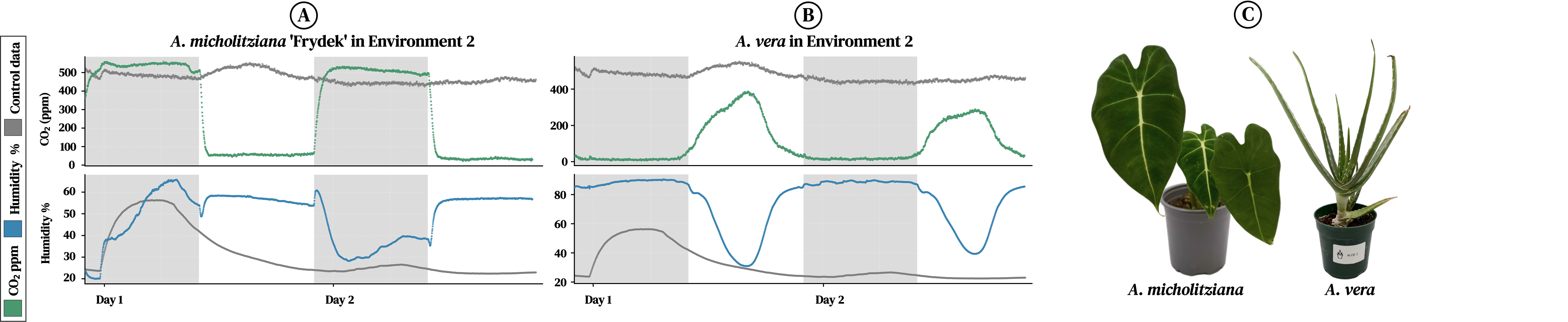}
  \caption{(A) Time series from \textit{A. micholitziana} (C\textsubscript{3}) showing dominant daytime CO\textsubscript{2} assimilation and no nocturnal uptake.  (B) Time series from \textit{A. vera} (obligate CAM) showing clear nocturnal CO\textsubscript{2} uptake and daytime stomatal closure consistent with CAM physiology. (C) Photographs of the two monitored specimens.}
  \Description{Three-panel figure showing photosynthetic patterns in two additional plant species. Panel A displays a line graph of CO2 concentration and humidity over time for A. micholitziana, a C3 plant, showing CO2 depletion during daytime with no nighttime uptake, characteristic of C3 photosynthesis. Panel B shows a corresponding graph for A. vera, an obligate CAM plant, displaying nocturnal CO2 uptake and minimal daytime gas exchange, consistent with CAM physiology. Panel C contains photographs of both plant specimens used in the measurements, showing the distinct morphologies of the plants.}
  \label{fig: other_species}
\end{figure}

To evaluate PhytoBits' generality across taxonomically and morphologically distinct plants, we monitored a C\textsubscript{3} species \textit{Alocasia micholitziana} and an obligate CAM species \textit{Aloe vera} concurrently and compared their diel gas-exchange signatures (Figure \ref{fig: other_species}A-C). Panel A shows shows \textit{A. micholitziana} with dominant daytime CO\textsubscript{2} assimilation and negligible nocturnal uptake, consistent with a C\textsubscript{3} strategy. Panel B shows  \textit{A. vera} time series with clear nocturnal CO\textsubscript{2} uptake and suppressed daytime assimilation, consistent with canonical CAM physiology

These recordings were done through another microcontroller - Micro:bit, and the data demonstrates that PhytoBits robustly resolves the expected temporal signatures of both CAM and C\textsubscript{3} photosynthesis across species with very different leaf form, water-use strategies, and stomatal behavior. Although absolute signal magnitude and fine-scale timing vary between species, likely reflecting differences in leaf morphology, stomatal density and conductance, the principal diel patterns (nocturnal fixation for CAM; daytime assimilation for C\textsubscript{3}) are unambiguous in each case. 

\begin{figure}[!h]
  \centering
  \includegraphics[width=0.95\linewidth]{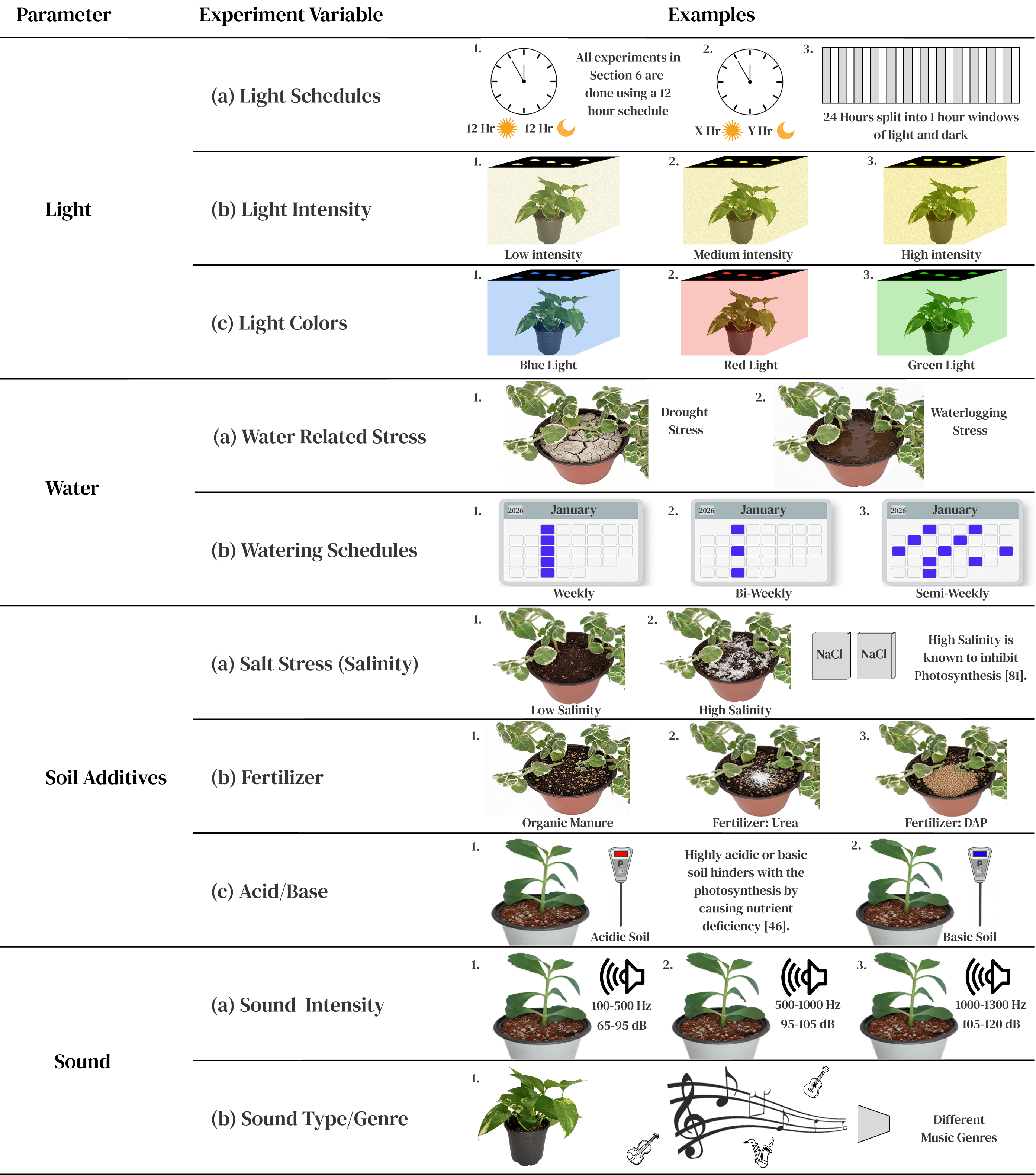}
  \caption{
  Design Parameters and Functional Capabilities of the PhytoBits Toolkit for Environmental Simulation.
  }
  \Description{Comprehensive table showing design parameters and experimental possibilities with the PhytoBits toolkit. The table is organized by environmental parameters in rows: Light (including options for schedules, intensity, and colors), Water (including irrigation schedules, water stress), Soil Additives (including fertilizers, salts, pH modifiers), and Sound. Each parameter row contains multiple example experiments demonstrating how students and researchers can creatively manipulate these variables to investigate plant physiological responses. The table illustrates the toolkit's versatility for designing diverse experimental conditions beyond basic photosynthesis measurements, enabling exploration of how different environmental factors affect plant gas exchange and CAM/C3 photosynthetic pathways.}
  \label{fig:design_parameters_table}
\end{figure}

\section{Real-world Application Scenarios}\label{section:real-world-applications}

\subsection{Extended Experimental Possibilities}
The Figure \ref{fig:design_parameters_table} shows some of the many environmental manipulations that can be tested with the PhytoBits toolkit to study photosynthesis. Light variables include intensity (all values for light intensity in our experiments reported in Section \ref{sec:env_setup}) and spectral colors such as blue, red, green, and other colors. Water treatments include drought by withholding water (see Sections \ref{facultative_CAM_section} and \ref{watering_regime_subsection}), water-logging which can inhibit photosynthesis \cite{waterlogging_stress}, and varied watering schedules (weekly, bi-weekly, semi-weekly). In our experiments we used a weekly watering schedule unless logging exceeded one week to avoid spurious temporal effects.

Chemical and edaphic factors include salinity gradients (salt stress is known to inhibit photosynthesis \cite{salt_stress}), fertilizer regimes, and soil pH (low pH may hinder photosynthesis \cite{ph_effect}). Different sound frequencies, intensities, and music genres may affect photosynthetic activity and carbon fixation \cite{sound_effect}. This can be tested and validated using the PhytoBits toolkit.

\subsection{Utility in Plant Physiology Research}
Our sensor measurements revealed clear differences in photosynthetic gas-exchange timing between well-watered and drought-stress conditions (Figure \ref{fig:facultative_CAM}).
Guided by these signatures, leaf tissue can be obtained during distinct physiological states, such as mixed diurnal-nocturnal uptake phases and nocturnal-only uptake phases, and conduct molecular analysis. For example, RNA sequencing can be performed using standard RNA-seq workflows, including RNA extraction, sequencing, and library preparation. These approaches enable quantification of expression levels for CAM-associated genes, such as \textit{PEPC, PPCK, PPDK, MDH} and \textit{RuBP}, allowing researchers to directly relate gene expression dynamics to in situ gas-exchange behavior \cite{kong2020molecular, yan2022transcriptome, heyduk2022differential}.

\subsection{Democratization for Community Engagement}
The toolkit's low cost, accessibility, and ease of assembly make it suitable not only for classroom use but also for deployment in the homes of plant hobbyists and community members. With the full tutorial and open-source materials provided, individuals can investigate any plant species and adapt the system to their own creative interests. This opens opportunities for informal science learning, community-driven experimentation, and broader public engagement with plant physiology.

When used in indoor environments, users should be mindful of ambient CO\textsubscript{2} levels. Poorly ventilated or tightly sealed rooms may accumulate elevated CO\textsubscript{2} concentrations, which can influence sensor readings and plant behavior. By monitoring both the plant and its surrounding environment, community users can gain a richer understanding of how microclimate conditions shape plant physiology.

\section{Discussion and Limitations}
PhytoBits occupied a productive middle ground between high-end physiological instruments and purely conceptual instruction for photosynthesis education. Although low-cost sensors lack the precision and control of LI-COR systems, their temporal patterns and phase relationships align closely enough with ground-truth measurements to support both authentic inquiry and introductory research.

\subsection{Semi-Sealed Pod Design and Experimental Controls}
A critical design feature is the intentionally semi-sealed leaf pod. During respiration periods (e.g., \textit{E. aureum} at night), CO\textsubscript{2} stabilizes around 500~ppm rather than rising indefinitely, reflecting minor leakage. Complete sealing would deprive plants of CO\textsubscript{2} for photosynthesis and deplete O\textsubscript{2} for respiration. This semi-sealed design maintains plant viability while enabling robust long-term monitoring.
Sealing methods can be adjusted based on context. Parafilm provides tighter seals for poorly ventilated spaces or high-occupancy rooms where ambient CO\textsubscript{2} fluctuations introduce artifacts. In well-ventilated environments, the simpler bag-tying method suffices. The semi-sealed environment creates measurable differences between enclosed and unenclosed leaves. The enclosed C\textsubscript{3} leaves experience  CO\textsubscript{2} limitation during photosynthesis, reducing rates compared to ambient air. This limitation is both an artifact and the detection mechanism. Flat CO\textsubscript{2} traces represent net balance among photosynthesis, respiration, and gas-exchange.
Good practice requires two controls: an empty leaf pod and an ambient sensor. If plant measurements resemble ambient controls, sealing is insufficient. If readings show flat lines divergent from controls, pod volume is too large relative to leaf area, users should add leaves or reduce pod volume.

\subsection{Quantitative Limitations.}
Validation in this study is based on visual alignment of CO\textsubscript{2} concentration time series (PhytoBits, units: ppm) with net assimilation rates (LI-COR LI-6400~XT, units: \(\mu\)mol~CO\textsubscript{2}~m\textsuperscript{$-$2}~s\textsuperscript{$-$1}). These two quantities are not dimensionally equivalent; the inverse correlation observed supports qualitative timing agreement but does not constitute a quantitative flux calibration. Per-chamber leak rates were not independently characterized in the current study, precluding conversion of concentration signals to absolute flux estimates.Formal Bland--Altman analysis and RMSE quantification are planned once leak-rate characterization is complete (see Section~\ref{sec:futurework}).

\subsection{Educational Applications and Accessibility}
PhytoBits is designed to make plant physiology education more interpretable and accessible. Its design is intended to invite students to explore trade-offs among enclosure tightness, sensor placement, and sampling rate, with the potential to support inquiry-based learning, data literacy, and interdisciplinary thinking. By providing a platform for students to test hypotheses about light schedules, watering, and temperature, the toolkit is designed to make metabolic plasticity observable in facultative species that transition between CAM and C\textsubscript{3}. It is also designed to support computational thinking through data cleaning, visualization, and debugging, and to lower the barrier to making core physiology concepts such as stomatal conductance, water-use efficiency, and circadian regulation empirically accessible.

PhytoBits costs approximately \$32--72 (Table \ref{tab:component_cost_table}), compared with roughly \$40{,}000--50{,}000 for LI-COR systems. This large cost difference substantially lowers barriers for under-resourced institutions and makes multi-site or distributed deployments more feasible than would be practical with commercial instrumentation.

\section{Future Work}
\label{sec:futurework}
Future work will pursue four directions. 
First, we will characterize the per-chamber leak through a CO\textsubscript{2} injection-decay calibration, enabling conversion of concentration offset signals into semi-quantitative leaf flux estimates and quantifying the reproducibility of the hand-assembled sealing method.
Second, analysis of existing control chamber data across multi-day light/dark cycles will formally bound sensor drift contributions to observed diurnal signals, alongside ambient CO₂ reference logging for signal normalization.
Third, we plan to extend the platform toward optional valve-assisted transient flux calibration, providing absolute flux anchoring while preserving the passive multi-day monitoring capability of the current design.
Finally, a classroom deployment study will assess learning outcomes, toolkit usability, and the pedagogical value of longitudinal plant physiology experiments across participants with varying technical backgrounds.

\section{Conclusion}
PhytoBits is a frugal toolkit that makes plant gas exchange visible and analyzable in classrooms and low-resource research settings. By integrating accessible materials to construct leaf pod, a low-cost CO\textsubscript{2} sensor, and a microcontroller suited for different programming, the system is designed to enable students and researchers to explore stomatal rhythms, photosynthesis strategies, and stress responses in plants. Validation against LI-COR instruments and biochemical assays suggested that the toolkit captured physiologically meaningful patterns.

\begin{acks}
Figures~\ref{fig:first_figure}, \ref{fig:toolkit_working_principle}, and \ref{fig:design_parameters_table} include AI-generated images.
This work was supported in part by the American Society of Plant Biologists (ASPB) Plant BLOOME grant, Northwestern University, and the Chicago Botanic Garden.
\end{acks}

\bibliographystyle{ACM-Reference-Format}
\bibliography{sample}

\appendix

\end{document}